\documentclass[a4paper, 12pt]{article}

\usepackage[utf8]{inputenc}
\usepackage[T1]{fontenc}
\usepackage[english]{babel}
\usepackage{lmodern} 
\usepackage{graphicx}
\usepackage{microtype}

\usepackage{bbm}
\usepackage{amsmath}
\usepackage{amsthm}
\usepackage{amssymb}
\usepackage{mathrsfs}
\usepackage[babel=true]{csquotes}
\usepackage{verbatim}
\usepackage{ytableau}

\usepackage{hyperref}

\definecolor{rougef}{rgb}{0.7,0,0}
\definecolor{vertf}{rgb}{0,0.6,0}
\definecolor{bleuf}{rgb}{0,0,0.9}

\newcommand{\diff}[1]{\text{d}#1\ }
\newcommand{\diffn}[2]{\text{d}^{#1}#2\ }
\newcommand{\diffx}{\text{d}}
\newcommand{\diffz}{\text{q}}
\newcommand{\diffhat}{\widehat{\rm{d}}}
\newcommand{\dZ}{\text{d}Z}
\newcommand{\dz}{\text{d}z}
\newcommand{\dzb}{\text{d}\zbar}
\newcommand{\dx}{\text{d}x}

\newcommand{\intyz}{\int \text{d}^2y\,\text{d}^2z\,}
\newcommand{\intybzb}{\int \text{d}^2\ybar\,\text{d}^2\zbar\,}

\newcommand{\cc}[1]{{#1}^{\ast}}
\DeclareMathOperator{\fdeg}{deg}
\newcommand{\transp}[1]{#1{}^{T}}
\newcommand{\inv}[1]{#1{}^{-1}}
\DeclareMathOperator{\idmat}{\mathbb{I}}
\DeclareMathOperator*{\lilsum}{{\displaystyle\Sigma}}

\newcommand{\zyx}[3]{\left(#3,#1; #2\right)}
\newcommand{\zzyyx}[5]{\left(#5,#1,#2;#3,#4\right)}
\newcommand{\yx}[2]{\left(#2;#1\right)}
\newcommand{\yyx}[3]{\left(#3;#1,#2\right)}

\newcommand\Bk[2]{\delta_{#1,#2}}

\DeclareMathOperator*{\starprod}{\bigstar}
\newcommand{\starpuis}[2]{\left(#1\right)^{\star #2}}
\newcommand{\starexp}[1]{\exp_{\star}\left(#1\right)}
\newcommand{\Pexp}[1]{P\exp_{\star}\left(#1\right)}

\newcommand{\starcomm}[2]{\left[#1,#2\right]_{\star}}
\newcommand{\staracomm}[2]{\left\{#1,#2\right\}_{\star}}

\newcommand{\alphadot}{\dot{\alpha}}
\newcommand{\betadot}{\dot{\beta}}
\newcommand{\gammadot}{\dot{\gamma}}

\newcommand{\underalpha}{\underline{\alpha}}
\newcommand{\underbeta}{\underline{\beta}}

\newcommand{\ibar}{\bar{\imath}}
\newcommand{\jbar}{\bar{\jmath}}

\newcommand{\nbar}{\bar{n}}
\newcommand{\ybar}{\bar{y}}
\newcommand{\zbar}{\bar{z}}

\newcommand{\pibar}{\bar{\pi}}

\newcommand{\lambdabar}{\bar\lambda}
\newcommand{\mubar}{\bar\mu}

\newcommand{\Mu}{M}

\newcommand{\fhat}{\hat{f}}
\newcommand{\ghat}{\hat{g}}

\newcommand{\WLczyx}[4]{W_{#1}(#4,#2;#3)}

\newcommand{\PAltemSA}[1]{{\mathcal{A}}_{n_0,t}\left(#1\vert\Mu\right)}
\newcommand{\PAffq}{g_{n_0}\left(\Lambda_i\right)}
\newcommand{\PAfftm}{\bar{f}_{n_0,t}\left(\lambdabar_i\vert \mubar\right)}
\newcommand{\PAffem}{f_{n_0,e}\left(\lambda_i\vert\mu\right)}
\newcommand{\PAffqSA}[2]{g_{n_0}\left(#1\vert#2\right)}
\newcommand{\PAfftmSA}[1]{\bar{f}_{n_0,t}\left(#1\vert \mubar\right)}
\newcommand{\PAffemSA}[1]{f_{n_0,e}\left(#1\vert\mu\right)}

\newcommand{\GYsv}[2]{x_{#1,#2}}
\newcommand{\GYisv}[2]{\check{x}_{#1,#2}}

\newcommand{\GYisvb}[2]{\check{\bar{x}}_{#1,#2}}
\newcommand{\GYosv}[1]{x_{0,#1}}
\newcommand{\GYoisv}[1]{\check{x}_{0,#1}}

\newcommand{\GYoisvb}[1]{\check{\bar{x}}_{0,#1}}
\newcommand{\GYpsu}[1]{{\chi}_{#1}}
\newcommand{\GYpsb}[1]{{\bar\chi}_{#1}}
\newcommand{\GYpnu}[1]{{\nu}_{#1}}
\newcommand{\GYpnb}[1]{{\bar\nu\,}_{#1}}
\newcommand{\GYbtbpf}[1]{K_{#1}}
\newcommand{\GYbtbY}[1]{\mathcal{K}_{#1}}
\newcommand{\GYbtbL}[1]{\mathcal{\tilde{K}}_{#1}}
\newcommand{\GYbtbLmb}[1]{\mathcal{\tilde{K}}^{MB}_{#1}}
\newcommand{\GYsau}[1]{\varepsilon_{#1}}
\newcommand{\GYsnO}[1]{\sigma_{#1}}
\newcommand{\GYciQ}[3]{Q_{#1}}
\newcommand{\GYciP}[2]{P_{#1,#2}}
\newcommand{\GYciPt}[2]{\tilde{P}_{#1,#2}}

\newcommand{\CSu}[1]{\Sigma_{#1}}
\newcommand{\CSb}[1]{{\bar\Sigma}_{#1}}
\newcommand{\CSdm}[2]{\det{}_{#1 , #2}}
\newcommand{\CSSnu}[1]{S_{#1}}
\newcommand{\CSSnb}[1]{\bar{S}_{#1}}
\newcommand{\CSSnd}[1]{\det(S_{#1})}
\newcommand{\CSSniu}[2]{S_{#1}^{(#2)}}
\newcommand{\CSSnib}[2]{\bar{S}_{#1}^{(#2)}}

\newcommand{\CFTpa}{\langle J_1,...,J_{n_0}\rangle_{\rm cyclic}}
\newcommand{\CFTwc}[2]{\langle #1#2\rangle}

\numberwithin{equation}{section}

\begin{document}

\begin{titlepage}

\setcounter{page}{1}

\begin{center}

\hfill


{\LARGE Noncommutative Wilson lines in higher-spin theory 
and correlation functions of conserved currents for free conformal fields}


\vskip 30pt

{\sc Roberto Bonezzi\,$^a$, Nicolas Boulanger\,$^{a}$, David De Filippi\,$^{a}$ \\{\small{and}} Per Sundell\,$^b$\\}

\vskip 30pt

{\em $^{a}$ \hskip -.1truecm Groupe de M\'ecanique et Gravitation,
Unit of Theoretical and Mathematical Physics,\\
University of Mons-- UMONS, 20 place du Parc, 7000 Mons, Belgium}

\vskip 10pt

{\em $^{b}$ \hskip -.1truecm Departamento de Ciencias F\'isicas, 
  Universidad Andres Bello,\\ Republica 220, Santiago de Chile}

\vskip 15pt

\end{center}

\vskip 30pt

\paragraph{Abstract.} 
We first prove that, in Vasiliev's theory, the zero-form 
charges studied in 
\href{https://arxiv.org/abs/1103.2360}{1103.2360} 
and 
\href{https://arxiv.org/abs/1208.3880}{1208.3880}
are twisted open Wilson lines in the noncommutative $Z$ space.
This is shown by mapping Vasiliev's higher-spin model on 
noncommutative Yang--Mills theory.
We then prove that, prior to Bose-symmetrising, 
the cyclically-symmetric  
higher-spin invariants given by the leading order of these 
$n$-point zero-form charges 
are equal to corresponding cyclically-invariant building blocks of 
$n$-point correlation functions of bilinear operators 
in free conformal field theories (CFT) in three dimensions.
On the higher spin gravity side, our computation reproduces the 
results of  
\href{https://arxiv.org/abs/1210.7963}{1210.7963}
using an alternative method
amenable to the computation of subleading corrections
obtained by perturbation theory in normal order.
On the free CFT side, our proof involves the explicit
computation of the separate cyclic building blocks of the  
correlation functions of $n$ conserved currents in 
arbitrary dimension $d>2\,$ using polarization vectors,
which is an original result. It is shown to agree, for
$d=3\,$, with the results obtained in \href{https://arxiv.org/abs/1301.3123}{1301.3123} in
various dimensions and where polarization spinors were used.

\vspace{1.5cm}
\begin{flushleft} \footnotesize
{${}^a$ \href{mailto:roberto.bonezzi@umons.ac.be}{roberto.bonezzi@umons.ac.be}, \href{mailto:nicolas.boulanger@umons.ac.be}{nicolas.boulanger@umons.ac.be}, \href{mailto:david.defilippi@umons.ac.be}{david.defilippi@umons.ac.be}}\\
{${}^b$ \href{mailto:per.anders.sundell@gmail.com}{per.anders.sundell@gmail.com}}\\

\end{flushleft}

\end{titlepage}


\newpage

{
\tableofcontents }

\section{Introduction}

The conjectured holographic duality between 
Higher Spin (HS) gauge theory on $AdS_{d+1}$ background and 
free $CFT_d$ was spelled out,  
after the pioneering works 
\cite{Flato:1978qz,Flato:1980zk,Bergshoeff:1988jm,Konstein:2000bi,Sundborg:2000wp,Sezgin:2001zs,Mikhailov:2002bp} 
in the references 
\cite{Sezgin:2002rt,Klebanov:2002ja,Leigh:2003gk,Sezgin:2003pt}, 
where the last work also contains a simple test of the conjecture. 
As stressed respectively in \cite{Sezgin:2001zs} and 
\cite{Klebanov:2002ja}, this holographic 
correspondence is remarkable in that it
(i) relates two weakly coupled theories and (ii)
does not require any supersymmetry.

If one postulates the validity of the HS/CFT conjecture, 
the holographic reconstruction programme undertaken in 
\cite{Bekaert:2014cea}, that starts from the free $O(N)$ 
model on the boundary, enabled the reconstruction 
of all the cubic vertices in the bulk \cite{Sleight:2016dba} 
as well as the quartic vertex for the bulk scalar field \cite{Bekaert:2015tva}.
The cubic vertices obtained in this way involve finitely many 
derivatives at finite values of the spins. 
On the other hand, without postulating the holographic 
conjecture, the standard tests of the $HS_4/CFT_3$ 
correspondence work well \cite{Giombi:2009wh,Giombi:2010vg} 
for the cubic vertices in Vasiliev's theory that are fully 
fixed by the rigid, nonabelian higher-spin symmetry 
algebra of the vacuum $AdS_4$ solution; for recent
results, see \cite{Sezgin:2017jgm,Didenko:2017lsn}.

The computations of \cite{Giombi:2010vg} of 3-point functions
show divergences in the case of the Vasiliev vertices 
that are not fully fixed by the pure higher-spin 
kinematics.
These divergences have been confirmed since then 
from a different perspective in \cite{Boulanger:2015ova},
where all the first nontrivial interactions around $AdS_4$ 
have been extracted from the Vasiliev equations.  
To date, no regularisation scheme has been shown to 
tame these divergences in a completely satisfactory
manner; for discussions, see for example \cite{Giombi:2012ms},
and for recent progress, see \cite{Sezgin:2017jgm,Didenko:2017lsn,Vasiliev:2017cae}.
Nonetheless. since the computation of \cite{Giombi:2010vg} utilises 
techniques of AdS/CFT that rely on the existence of a standard 
action principle on the bulk side, i.e. an action for 
which the kinetic terms are given by Fronsdal Lagrangians
in $AdS_4$ \cite{Fronsdal:1978rb}, 
it is suggestive of the existence of 
an effective action of deformed Fronsdal type, but not
necessarily of any underlying path integral measure for 
self-interacting Fronsdal fields.

Being more circumspect, 
one can say that, for the Vasiliev's equations, 
the issues of locality and weak-field 
perturbative expansion around $AdS_4$ 
are subtle and require more investigations, 
see \cite{Skvortsov:2015lja,Taronna:2017jeq,Vasiliev:2016xui,Sleight:2017pcz,Didenko:2017lsn, Gelfond:2017wrh, Misuna:2017bjb, Vasiliev:2017cae} 
for some preliminary works in that direction. 
In fact, taking a closer look at the deformed
Fronsdal action, it is more reminiscent of an effective
than a classical action.
The weakest assumption would be a duality between Vasiliev’s equations 
and the deformed Fronsdal theory, in the sense that the two 
theories would be equivalent only at the level of their 
respective on-shell actions subject to suitable boundary conditions. 
However, in view of the matching of 3-point couplings 
\cite{Sezgin:2003pt,Giombi:2010vg,Sezgin:2017jgm,Didenko:2017lsn}, 
a more natural outcome would be that Vasiliev’s equations 
contain a deformed Fronsdal branch, 
obtained by a fine-tuned field redefinition \cite{Vasiliev:2016xui} 
possibly related to a noncommutative geometric 
framework \cite{Iazeolla:2011cb,Boulanger:2015kfa,Iazeolla:2017vng}.
As this branch would have to correspond to a unitary 
(free) CFT, one should thus think of the deformed Fronsdal 
action as an effective action, that is, there is no need 
to quantise it any further --- whereas further $1/N$ corrections can be 
obtained by altering boundary conditions.
Moreover, unlike in an ordinary quantum field theory, 
in which the effective action has a loop expansion, 
the trivial nature of the $1/N$ expansion of the
free conformal field theory implies that the anti-holographically dual
deformed Fronsdal action has a trivial loop expansion as well.
Thus, rather than to quantise the deformed Fronsdal theory
as a four-dimensional field theory (only to discover that all loop 
corrections actually cancel), it seems more reasonable to us
that the actual microscopic theory turns to be a 
field theory based on a classical action of the 
covariant Hamiltonian form proposed in \cite{Boulanger:2015kfa}.

In this paper, we shall examine the issue of locality 
of unfolded field equations from a different point of view,
by studying higher spin invariants known 
as \emph{zero-form charges}  \cite{Engquist:2005yt,Sezgin:2005pv}.
It was argued in \cite{Boulanger:2008tg,Colombo:2010fu} that 
rather than asking for locality in a gravitational gauge theory at the
level of the field equations, the question is to determine 
in which way the degrees of freedom assemble themselves 
so as to exhibit some form of cluster decomposition, in 
a way reminiscent to glueball formation in the strong coupling regime
of Quantum Chromo Dynamics (QCD). 
In the body of this paper, we shall present further
evidence for that this heuristic comparison with QCD 
is not coincidental.

In Vasiliev's unfolded formulation,
where the spacetime dependence of the 
locally-defined master fields can be encoded
into a gauge function, the boundary condition
dictated by the space-time physics that one wishes 
to address is translated into the choice of a suitable class
of functions in the internal fiber space.
Following this line of reasoning, the authors of 
\cite{Colombo:2010fu} proposed that locality 
properties manifest themselves at the level of zero-form charges
\cite{Sezgin:2011hq}, whose leading orders
in the expansion in terms of curvatures of the bulk-to-boundary propagators 
found in \cite{Giombi:2010vg} hence ought to reproduce holographic 
correlation functions of the dual $CFT_3\,$
as was later verified in \cite{Colombo:2012jx} for two- and three-point functions.
A related check was performed in the case of black-hole-like solutions 
in \cite{Iazeolla:2011cb,Iazeolla:2012nf,Iazeolla:2017vng}, whose asymptotic charges are mapped 
to non-polynomial functions in the fiber space, that in 
their turn determine zero-form charges that indeed exhibit 
cluster-decomposition properties, whereby the zero-form
charges of two well-separated one-body solutions
are perturbatively additive in the leading order
of the separation parameter.

The relation between the leading orders of more general zero-form 
charges and general $n$-point functions was then established 
in \cite{Didenko:2012tv} following a slightly different approach 
in terms of Cayley transforms, reproducing the $CFT_3$ results for 
the 3-point correlation functions of conserved currents for free bosons 
and free fermions obtained in \cite{Giombi:2011rz,Maldacena:2011jn}.
In \cite{Didenko:2013bj}, the $n$-point 
functions involving the conformal weight $\Delta=2$
operator 
were also computed and a comparison was made 
between the leading-order zero-form charges 
of \cite{Colombo:2012jx}
and the results of \cite{Gelfond:2013xt}, 
showing complete agreement. The authors of 
\cite{Gelfond:2013xt} used a convenient twistor basis 
in order to express the operator product algebra of free
bosons and fermions in various dimensions, and from this 
they computed all the correlators. The case of free $CFT_3$
was fully covered, as well as free $CFT_4$ (all Lorentz spins), 
upon certain truncations of the $sp(8)$ generalised spacetime 
coordinates. No expressions were given, however, for the $n$-point 
correlation functions in the free scalar $CFT_d$ for arbitrary 
$d\,$. The latter were conjectured in \cite{Sleight:2016dba}. 

We want to stress that the approach to the computation 
of free CFT correlation functions initiated in \cite{Colombo:2012jx} from the bulk side  
has a priori nothing to do with the usual AdS/CFT 
prescription \cite{Gubser:1998bc,Witten:1998qj}.
From the standard approach in terms of Witten diagrams, 
it is rather surprising that the evaluation of some quantities 
in the bulk (here the zero-form charges), 
evaluated on the free theory, could produce $n$-point 
correlation functions on the CFT side, 
as no information from vertices of order $n$
in the Vasiliev theory is being used.

It can  be somehow reconciled with holography if one uses a 
non-standard HS action \cite{Boulanger:2012bj}
that reproduces the Vasiliev equations upon variation and
is non-standard in the sense that the Lagrangian density 
is integrated over a higher-dimensional noncommutative open manifold containing $AdS_4$ as 
a submanifold of its boundary,  and with kinetic 
terms that are \emph{not} of the Fronsdal type 
but instead of the $BF$ type usually met in 
topological field theory.
In this approach,  the zero-form charges appear as 
some pieces of the on-shell action. 
The fact that 
the generating functional in the HS bulk theory 
reproduces, in the semi-classical limit, 
the free CFT generating function of correlation 
functions, 
is therefore not totally odd \cite{Colombo:2012jx}.

In the present paper, one of the tasks we undertake 
is to pursue the evaluation of zero-form charges 
along these lines, using throughout a method that enables 
one to evaluate subleading corrections to the 
free CFT correlators due to the higher
orders in the weak field expansion of Vasiliev's  zero-form 
master field, postponing the systematic evaluation of these 
subleading corrections to a future work. 
Although the results of \cite{Didenko:2012tv,Didenko:2013bj}
already produced the $n$-point correlators for free bosons
and fermions $CFT_3\,$, the fact that the noncommutative 
$Z$ variables at the heart of Vasiliev's formalism \cite{Vasiliev:1992av} were discarded ab initio does not allow the consideration of subleading 
corrections in Vasiliev's equations.

The plan of the paper is a follows: After a review 
of Vasiliev's bosonic model in Section \ref{sec:Vasiliev}, 
Section \ref{sec:Obs} contains a description of  
observables in Vasiliev's model and the proof
(completed in Appendix \ref{sec:thmWL}) that the zero-form 
charges discussed in \cite{Colombo:2010fu,Sezgin:2011hq,Colombo:2012jx}, 
are nothing but twisted open Wilson lines in the 
noncommutative twistor $Z$ space.
In Section \ref{sec:Correlators}, we compute the 
twisted open Wilson lines on the bulk-to-boundary 
propagator computed in \cite{Giombi:2010vg} and derive 
the corresponding quasi-amplitudes for arbitrary 
number of external legs. 
In the next Section \ref{sec:CFT}, we compute the 
$n$-point correlation functions of conserved currents of the
free $CFT$ corresponding to a set of free bosons 
in arbitrary dimension $d\,$. 
We show that, even before Bose symmetrisation, 
the cyclic-invariant pre-amplitudes 
obtained from the open Wilson lines 
at leading order in weak field expansion correspond 
to the cyclically-invariant building blocks 
of the correlators in the free $U(N)$ model,  
obtained from the Wick contraction of the nearest-neighbours
free fields inside the correlation functions.
Our notation for spinor indices together 
with some technical results are contained 
in Appendix \ref{app:CS}, 
while we relegated some other 
technical results in the 
Appendix \ref{app:gauss}.

\section{Vasiliev's bosonic model}
\label{sec:Vasiliev}

Four-dimensional bosonic Vasiliev's higher spin theories are formulated in terms of locally 
defined differential forms on a base manifold  
$\mathcal{X}_4\times\mathcal{Z}_4\,$, 
where $\mathcal{X}_4$ is a commutative spacetime manifold, with coordinates $x^\mu$, and $\mathcal{Z}_4$ is a noncommutative four-manifold coordinatised by $Z^{\underalpha}$, with $\underalpha=1,..,4\,$. 
The fields are valued in an associative algebra generated by oscillators $Y^{\underalpha}$ that are coordinates of a noncommutative internal manifold $\mathcal{Y}_4\,$. 
By using symbol calculus, we shall treat $Z^{\underalpha}$ and $Y^{\underalpha}$ as commuting variables, whereby the noncommutative structure is ensured by endowing the algebra of functions\footnote{We put hats  
on objects that are nontrivial differential forms on $\mathcal{Z}_4\,$.} $\hat f(x,Z;Y)$ 
with a noncommutative associative product, denoted by $\star\,$, giving rise to the 
following oscillator algebra:
\begin{equation}
\starcomm{Y^{\underalpha}}{Y^{\underbeta}} = 2i\,C^{\underalpha\underbeta}
\;,\quad
\starcomm{Z^{\underalpha}}{Z^{\underbeta}} = -2i\,C^{\underalpha\underbeta}
\;,\quad
\starcomm{Y^{\underalpha}}{Z^{\underbeta}} = 0\;,
\end{equation}
where $C^{\underalpha\underbeta}$ is an $Sp(4)$ invariant non-degenerate tensor
used to raise and lower $Sp(4)$ indices using the NW-SE convention
\begin{equation}
V^{\underalpha}:=C^{\underalpha\underbeta}\,V_{\underbeta}
\;,
\quad V_{\underalpha}:=V^{\underbeta}\,C_{\underbeta\underalpha}
\;.     
\end{equation}
The $Sp(4)$ indices are split under $sl(2,\mathbb{C})$ in the following way:
\begin{equation}
Y^{\underalpha} = (y^\alpha,\ybar^{\alphadot})
\;,\quad
Z^{\underalpha} = (z^\alpha,-\zbar^{\alphadot})
\;.
\end{equation}
Correspondingly, the $Sp(4)$ invariant tensor is chosen to be 
\begin{equation}\label{eq:epsym}
C^{\underalpha\underbeta}
=
\begin{pmatrix} \epsilon^{\alpha\beta} & 0 \\ 0 & \epsilon^{\alphadot\betadot} \end{pmatrix}
\;,\quad
C_{\underalpha\underbeta}
=
\begin{pmatrix} \epsilon_{\alpha\beta} & 0 \\ 0 & \epsilon_{\alphadot\betadot} \end{pmatrix}
\;,
\end{equation}
where $\epsilon_{12}=\epsilon^{12}=1$ and $\epsilon_{\dot1\dot2}=\epsilon^{\dot1\dot2}=1$ for the $sl(2,\mathbb{C})$ invariant tensors.

The star product algebra is extended from functions to differential forms on $\mathcal{X}_4\times\mathcal{Z}_4\,$ by defining
\begin{equation}
\dZ^{\underalpha}\star\fhat:=\dZ^{\underalpha}\,\fhat
\;,\quad
\fhat\star\dZ^{\underalpha}:=\fhat\,\dZ^{\underalpha}
\;,
\end{equation}
\emph{idem} ${\rm d}x^\mu$, where $\fhat$ is a differential form 
and the wedge product is left implicit.
Thus, the differential forms are horizontal on a total
space $\mathcal{X}_4\times\mathcal{Z}_4\times {\cal Y}_4$, 
sometimes referred to as the correspondence space, 
with fiber space ${\cal Y}_4$, base $\mathcal{X}_4\times\mathcal{Z}_4$
and total horizontal differential 
\begin{equation}
\diffhat := \diffx + \diffz
\;,\quad
\diffx := \diffx x^\mu \partial^x_{\mu}
\;,\quad
\diffz := \dZ^{\underalpha} \partial^Z_{\underalpha}
\;,
\end{equation}
which obeys the graded Leibniz rule
\begin{equation}
\diffhat\left(\fhat\star\ghat\right)
=
\left(\diffhat\fhat\right)\star\ghat
+(-)^{\fdeg\fhat\fdeg\ghat}
\fhat\star\left(\diffhat\ghat\right)
\;,
\end{equation}
with deg denoting the total form degree.
The differential graded star product algebra of forms 
admits a set of linear (anti-)automorphisms defined by 
\begin{align}
\label{eq:def pi}
\pi(x,y,\ybar,z,-\zbar) &= (x,-y,\ybar,-z,-\zbar)
\;,\\
\pibar(x,y,\ybar,z,-\zbar) &= (x,y,-\ybar,z,\zbar)
\;,\\
\tau(x,y,\ybar,z,-\zbar) &= (x,iy,i\ybar,-iz,i\zbar)
\;,
\end{align}
and 
\begin{align}
&
\pi(\diffhat\fhat)=\diffhat\,\pi(\fhat)
\;,\quad
\pi(\fhat\star\ghat)=\pi(\fhat)\star\pi(\ghat)
\;,\quad
\text{idem for}\;\pibar
\;,\\&
\tau(\diffhat\fhat)=\diffhat\,\tau(\fhat)
\;,
\quad \tau(\fhat\star\ghat)=
(-)^{\fdeg\fhat\fdeg\ghat}\tau(\ghat)\star\tau(\fhat)
\;,
\end{align}
for differential forms $\fhat$ and $\ghat\,$. Let us notice that $\tau^2=\pi\pibar$ and demanding that $\pi\pibar(\fhat)=\fhat$ amounts to removing all half-integer spin 
gauge fields on ${\cal X}_4$ from the model, leaving a bosonic model whose
gauge fields on ${\cal X}_4$ have integer spin.
The hermitian conjugation is the anti-linear anti-automorphism defined by 
\begin{equation}
(x,y,\ybar,z,-\zbar)^\dagger = (x,\ybar,y,\zbar,-z)
\;,
\end{equation}
\begin{equation}
(\diffhat\fhat)^\dagger=\diffhat\fhat^\dagger
\;,\quad
(\fhat\star\ghat)^\dagger=(-)^{\fdeg\fhat\fdeg\ghat}
\,
\ghat^\dagger\star\fhat^\dagger
\;.
\end{equation}

In what follows, we shall use the normal ordered basis for the star product\footnote{
It provides normal ordering of 
$a^{-}_{\underalpha} :=\tfrac{1}{2}(Y_{\underalpha}+Z_{\underalpha})\,$ and 
$a^{+ \underalpha } :=\tfrac{1}{2i}(Y^{\underalpha}-Z^{\underalpha})\,$  
so that $\fhat\star a^-_{\underalpha} = \fhat\,a^-_{\underalpha}$ and 
$a^{+ \underalpha}\star\fhat=a^{+ \underalpha}\,\fhat\,$.
}.
Among the various conventions existing in the literature, 
we choose to work with the explicit realisation:
\begin{equation}
\big(\fhat\star\ghat\big)(Z,Y):=\int\frac{d^4Ud^4V}{(2\pi)^4}\,e^{iV^{\underalpha}U_{\underalpha}}\,\fhat(Z+U,Y+U)\,\ghat(Z-V,Y+V)
\;,
\end{equation}
for auxiliary variables $U^{\underalpha}:=(u^\alpha,\bar u^{\alphadot})$ 
and $V^{\underalpha}:=(v^\alpha,\bar v^{\alphadot})\,$.
The space of bounded functions of $Y$ and $Z$ whose complex modulus is integrable (usually written $L^1(\mathcal{Y}_4\times {\cal Z}_4)$)
forms a star product algebra that admits a trace operation, given by
\begin{align}
{\rm Tr}\,\hat 
f(Z,Y):=\int\diffn{4}{Z}\diffn{4}{Y}
\hat f(Z,Y)
\;;
\label{Trace}
\end{align}
indeed, it has the desired cyclicity property
\begin{align}
{\rm Tr}\,
\fhat \star \ghat =
{\rm Tr}\,\ghat \star \fhat
\;.
\end{align}
It has the remarkable property:
\begin{equation}
\label{eq:fact trace}
{\rm Tr}\,(\fhat(Y)\star\ghat(Z))
=
{\rm Tr}\,(\fhat(Y)\ghat(Z))\;.
\end{equation}

In normal order, the inner Klein operators ${\hat\kappa}$ and ${\hat{\bar\kappa}}$, defined by
\begin{equation}
\label{eq:pisfromks}
{\hat\kappa}\star {\hat\kappa}=1={\hat{\bar\kappa}}\star {\hat{\bar\kappa}}
\;,\quad
\pi(\fhat) = {\hat\kappa}\star \fhat\star {\hat\kappa}
\;,\quad
\pibar(\fhat) = {\hat{\bar\kappa}}\star \fhat\star {\hat{\bar\kappa}}
\;,
\end{equation}
for all zero-forms $\fhat$, 
become real-analytic functions on ${\cal Y}_4\times {\cal Z}_4$, 
\emph{viz.}
\begin{equation}
{\hat\kappa} := e^{iy^\alpha z_\alpha}
\;,\quad
{\hat{\bar\kappa}} := e^{-i\ybar^{\alphadot} \zbar_{\alphadot}}
\;.
\end{equation}
As shown in \cite{Didenko:2009td}, the inner kleinians factorise as
\begin{equation}
\begin{split}
\label{eq:k=ky*kz}
&{\hat\kappa} = \kappa_y(y)\star\kappa_z(z)
\;,\quad
{\hat{\bar\kappa}}={\bar\kappa}_y(\ybar)\star{\bar\kappa}_z(\zbar)
\;,\quad
{\rm with}
\\&
\kappa_y(y):=2\pi\,\delta^2(y)
\;,\quad
\kappa_z(z):=2\pi\,\delta^2(z)
\;,\quad
{\bar\kappa}_y(\ybar):=2\pi\,\delta^2(\ybar)
\;,\quad
{\bar\kappa}_z(\zbar):=2\pi\,\delta^2(\zbar)
\;,
\end{split}
\end{equation}
which implies that their symbols are non-real analytic in Weyl order.
As we shall see, the inner kleinians define closed
and central elements appearing in the equations of motion,
which implies that the fields will be real-analytic on-shell 
in normal order but not in Weyl order. Thus, the
normal order is suitable for a standard higher spin 
gravity interpretation of the model, which requires
a manifestly Lorentz covariant symbol calculus with
symbols that are real analytic at the origin of $\mathcal{Z}_4\times {\cal Y}_4\,$.

The master fields describing the bosonic model consist of a connection one-form 
$\widehat A$ and a twisted-adjoint zero-form $\widehat\Phi\,$, subject to the 
reality conditions and bosonic projection
\begin{align}
\widehat{A}^\dagger = -\widehat{A}
\;,\quad
\pi(\widehat{A}) &= \pibar(\widehat{A})
\;,\\
\label{eq:RC tafV}
\widehat{\Phi}^\dagger = \pi(\widehat{\Phi}) &= \pibar({\widehat{\Phi}})
\;.
\end{align}
At the linearised level the physical spectrum consists of an infinite tower of massless particles of every integer spin, each occurring once.
The bosonic model can be further projected to its minimal version, containing only even spin particles, by imposing
\begin{equation}
\label{eq:MBP tafV}
\tau(\widehat{A}) = -\widehat{A}\;,\quad\tau(\widehat{\Phi}) = \pi({\widehat{\Phi}})
\;.
\end{equation}
The equations of motion are given by the twisted-adjoint covariant constancy condition
\begin{equation}
\diffhat\widehat{\Phi} + \widehat{A}\star\widehat{\Phi} -\widehat{\Phi}\star\pi(\widehat{A}) = 0\label{eq:Vas2}
\;,
\end{equation}
and the curvature constraint
\begin{equation}
\diffhat\widehat{A} + \widehat{A}\star\widehat{A} +\widehat{\Phi}\star\widehat{J} = 0\label{eq:Vas1}
\;,
\end{equation}
where the two-form $\widehat{J}$ is given by 
\begin{equation}
\label{eq:defJ}
\widehat{J} := -\frac{i}{4}\big(e^{i\theta}\,{\hat\kappa}\,\dz^{\alpha}\dz_{\alpha} + e^{-i\theta}\,{\hat{\bar\kappa}}\,\dzb^{\alphadot}\dzb_{\alphadot}\big)
\;,
\end{equation}
with $\theta$ a real constant caracterising the model. 
This element obeys  
\begin{equation}
\widehat{J}^\dagger=\tau(\widehat{J})=-\widehat{J}\;,
\end{equation}
and
\begin{equation}
\begin{split}
&
\diffhat\widehat{J}\equiv0\;,\quad \widehat{A}\star\widehat{J}-\widehat{J}\star\pi(\widehat{A})\equiv0\equiv \widehat{\Phi}\star\widehat{J}-\widehat{J}\star\pi(\widehat{\Phi})
\;.
\end{split}
\end{equation}
It follows that the field equations are compatible with
the reality conditions on the master fields and the
integer-spin projections, and that they are universally 
Cartan integrable.
The latter implies invariance of (\ref{eq:Vas1}, \ref{eq:Vas2}) 
under the finite gauge transformation
\begin{align}
\label{eq:GIf cV}
\widehat{A} &\longrightarrow \widehat{g}\star\diffhat\widehat{g}^{-1} + \widehat{g}\star\widehat{A}\star\widehat{g}^{-1}
\;,\\
\label{eq:GIf tafV}
\widehat{\Phi} &\longrightarrow \widehat{g}\star\widehat{\Phi}\star\pi(\widehat{g}^{-1})
\;,\quad
\widehat{g} = \widehat{g}\zyx{Z}{Y}{x}
\;, 
\end{align}
which preserve the reality conditions on the master fields
and the integer-spin projection of the model provided that
\begin{equation}
\label{eq:RC gfV}
\widehat{g}^{\dagger} = \widehat{g}^{-1}
\;,\quad
\pi(\widehat{g})=\pibar(\widehat{g})
\;;
\end{equation}
the minimal projection in addition requires that 
\begin{equation}
\tau(\widehat{g}) = \widehat{g}^{-1}
\;.
\end{equation}
The parity invariant models \cite{Sezgin:2003pt} are obtained by taking 
$\theta=0$ for the Type A model and $\theta=\tfrac{\pi}{2}$ for the Type B model,
in which the physical scalar field is parity even and odd, respectively.
%

Upon splitting the connection one-form into 
$\diffx x^\mu$ and $\dZ^{\underalpha}$ directions: 
\begin{equation}
\widehat{A} = \widehat{U} + \widehat{V}
\;,\quad
\widehat{U} = \diffx x^\mu \widehat{A}_\mu
\;,\quad 
\widehat{V} = \dz^\alpha \widehat{A}_\alpha + \dzb^{\alphadot}\widehat{A}_{\alphadot}
\;,
\end{equation}
Vasiliev's equations read
\begin{align}
\label{eq:Vas1x}
\diffx\widehat{U} + \widehat{U}\star\widehat{U} &= 0
\;,\\
\label{eq:Vas1xz}
\diffx\widehat{V} + \diffz\widehat{U} + \widehat{U}\star\widehat{V} + \widehat{V}\star\widehat{U} &= 0
\;,\\
\label{eq:Vas1z}
\diffz\widehat{V} +\widehat{V}\star\widehat{V} + \widehat{\Phi}\star\widehat{J} &= 0
\;,\\
\label{eq:Vas2x}
\diffx\widehat{\Phi} + \widehat{U}\star\widehat{\Phi} - \widehat{\Phi}\star\pi(\widehat{U}) &= 0
\;,\\
\label{eq:Vas2z}
\diffz\widehat{\Phi} + \widehat{V}\star\widehat{\Phi} - \widehat{\Phi}\star\pi(\widehat{V}) &= 0
\;.
\end{align}
Remarkably, unlike the case of a connection on a 
commutative manifold,
the connection on a noncommutative (symplectic) manifold can be
mapped in a one-to-one fashion to an adjoint quantity, given
in Vasiliev's theory by\footnote{
The transformation of $\widehat{S}_{\underalpha}$ can be obtained 
by recalling that $-2i \partial^Z_{\underalpha}\hat{f}= 
[Z_{\underalpha}, \hat{f}]_\star\,$.
}
\begin{align}
\label{eq:def S,Psi}
\widehat{S}_{\underalpha} :=
Z_{\underalpha} - 2i\,\widehat{A}_{\underalpha} 
\;,\quad
\widehat\Psi := \widehat{\Phi}\star \hat\kappa
\;,\quad 
\widehat{\bar\Psi} := \widehat{\Phi}\star 
\hat{\bar\kappa}
\;.
\end{align}
In terms of these variables, Vasiliev's equations read as follows: 
\begin{align}
\label{EVS1}
\diffx\widehat{U} + \widehat{U}\star\widehat{U} 
&=0
\;,&
\starcomm{\widehat{S}_{\alpha}}{\widehat{S}_{\alphadot}}
&=0
\;,\\
\diffx\widehat{S}_{\alpha}+\starcomm{\widehat{U}}{\widehat{S}_{\alpha}}
&=0
\;,&
\diffx\widehat{S}_{\alphadot}+\starcomm{\widehat{U}}{\widehat{S}_{\alphadot}}
&=0
\;,\\
\diffx\widehat\Psi+\starcomm{\widehat{U}}{\widehat\Psi}
&=0
\;,&
\diffx\widehat{\bar\Psi}+\starcomm{\widehat{U}}{\widehat{\bar\Psi}}
&=0
\;,\\
\label{EVS4}
\staracomm{\widehat{S}_{\alpha}}{\widehat\Psi}
=
\starcomm{\widehat{S}_{\alpha}}{\widehat{\bar\Psi}}
&=0
\;,&
\staracomm{\widehat{S}_{\alpha}}{\widehat\Psi}
=
\starcomm{\widehat{S}_{\alpha}}{\widehat{\bar\Psi}}
&=0
\;,\\
\label{EVS5}
\starcomm{\widehat{S}_{\alpha}}{\widehat{S}_{\beta}}
+2i\epsilon_{\alpha\beta}
(1-e^{i\theta}\widehat\Psi)
&=0
\;,&
\starcomm{\widehat{S}_{\alphadot}}{\widehat{S}_{\betadot}}
+2i\epsilon_{\alphadot\betadot}
(1-e^{-i\theta}\widehat{\bar\Psi})
&=0
\;.
\end{align}
Thus, the adjoint variables $(\widehat{S}_{\alpha},\widehat{S}_{\alphadot})$
obey a generalized version of Wigner's deformation \cite{Wigner:50,Yang:51} of the Heisenberg algebra, 
as in \cite{Vasiliev:1989re}, 
and are hence referred to as the deformed oscillators.

The vacuum solution describing the $AdS_4$ background is obtained by setting $\widehat{\Phi}=0=\widehat{V}$ and taking $\widehat{U}=\Omega$, the Cartan connection of $AdS_4\,$, given by
\begin{equation}
\Omega(Y\vert x)
=\frac{1}{4i}\,
\big(y^\alpha y^\beta\,\omega_{\alpha\beta}
+\ybar^{\alphadot}\ybar^{\betadot}\,\bar{\omega}_{\alphadot\betadot}
+2\,y^\alpha\ybar^{\alphadot}\,h_{\alpha\alphadot}\big)
\;
\end{equation}
obeying the zero-curvature condition $\diffx\Omega+\Omega\star\Omega=0\,$.
One may then perform a perturbative expansion around this background and find, 
at the linearised level, the Central On Mass-Shell Theorem \cite{Vasiliev:1988sa} 
that describes, in a suitable gauge, 
the free propagation of an infinite tower of Fronsdal fields
around $AdS_4\,$.
For our purpose it is important to recall that the twisted adjoint zero-form is $Z$-independent at the linearised level, and will be denoted by $\Phi\yx{Y}{x}$.

\section{Observables in Vasiliev's theory}
\label{sec:Obs}

As the gauge transformations of Vasiliev's theory resemble 
those of noncommutative Yang-Mills theory (see \cite{Gross:2000ba, Ambjorn:2000cs,Makeenko:2002uj}), some results of the latter theory
may be applied to Vasiliev's theory. 
In particular, one can construct gauge invariant observables from holonomies formed from curves that are closed in ${\cal X}_4$ 
and open in ${\cal Z}_4$.
To this end, we consider a curve 
\begin{equation}
\label{eq:defCurve}
\mathcal{C} : 
[0,1] \rightarrow \mathcal{X}_4\times \mathcal{Z}_4 : 
\sigma \rightarrow (\xi^\mu(\sigma),\xi^{\underalpha}(\sigma))
\end{equation}
that is based at the origin, closed in the commutative directions
and open along the noncommutative space, 
\emph{i.e.}
\begin{equation}
\xi^\mu(0) = \xi^\mu(1) = 0
\;,\quad 
\xi^{\underalpha}(0)=0
\;,\quad
\xi^{\underalpha}(1)=2\Mu^{\underalpha}
=2\,C^{\underalpha\underbeta}\Mu_{\underbeta}
\;.
\end{equation}
Here $\Mu_{\underalpha}=(\mu_{\alpha},\mubar_{\alphadot})$ is seen as a momentum conjugated to $Z^{\underalpha}$.
We can associate the following Wilson line to $\mathcal{C}\,$:
\begin{align}
\WLczyx{\mathcal{C}}{Z}{Y}{x} 
:&=
\Pexp{\int_0^1 \diff\sigma 
\left(
\dot\xi^\mu(\sigma) \widehat{A}_\mu(\sigma)
+\dot\xi^{\underalpha}(\sigma) \widehat{A}_{\underalpha}(\sigma)
\right)}
\nonumber\\&=
\sum_{n=0}^{\infty} 
\int_{0}^1 \diff{\sigma_n} ... \int_{0}^{\sigma_{2}}\diff{\sigma_1}
\starprod_{i=1}^{n}\left(
\dot\xi^\mu(\sigma_i) \widehat{A}_\mu(\sigma_i)
+\dot\xi^{\underalpha}(\sigma_i) \widehat{A}_{\underalpha}(\sigma_i)
\right)
\;.
\end{align}
where $\widehat{A}(\sigma):=\widehat{A}\zyx{Z^{\underalpha}+\xi^{\underalpha}(\sigma)}{Y^{\underalpha}}{x^{\mu}+\xi^{\mu}(\sigma)}\,$
and
for any set of $n$ functions $\left\{\fhat_1,...,\fhat_{n}\right\}$,
the symbol $\starprod_{i=1}^{n}\fhat_i$ is defined as $\fhat_1\star...\star\fhat_{n}$ in that order.
Under the gauge transformation 
(\ref{eq:GIf cV}, \ref{eq:GIf tafV}) 
it transforms as :
\begin{equation}
\label{eq:WL}
\WLczyx{\mathcal{C}}{Z}{Y}{x} 
\longrightarrow 
\widehat{g}\zyx{Z}{Y}{x}\star\WLczyx{\mathcal{C}}{Z}{Y}{x}\star\widehat{g}^{-1}\zyx{Z+2\Mu}{Y}{x}
\;.
\end{equation}
Unlike open Wilson lines in commutative spaces, 
which cannot be made gauge invariant, their counterparts 
in noncommutative spaces can be made gauge invariant 
by star multiplying them with the generator $e^{i\Mu Z}$
of finite translations in ${\cal Z}_4\,$,
and tracing over both the fiber space and base manifold
\footnote{Let us point out that this $\star$-multiplication 
by $e^{i\Mu Z}$ is not equivalent to considering the same 
curve, closed by a straight path 
linking its two ends, since the connection $\widehat{A}$ is not 
path integrated along this straight path.}
.
Thus, for any adjoint zero-form $\hat{O}\zyx{Z}{Y}{x}$, \emph{i.e.} any 
operator transforming as 
$\hat{O}\longrightarrow\widehat{g}\star \hat{O}\star\widehat{g}^{-1}$, the 
quantity
\begin{equation}
\label{eq:Observable}
{\widetilde{O}}_{\mathcal{C}}\left(\Mu\vert x\right)
:=
{\rm Tr}\,\left[ \hat{O}\zyx{Z}{Y}{x} \star 
\WLczyx{\mathcal{C}}{Z}{Y}{x}
\star e^{i\Mu Z}\right]
\;,
\end{equation}
which one may formally think of as the Fourier transform of the impurity,
is gauge invariant, as shown in \cite{Makeenko:2002uj}.
Indeed, this follows from the cyclicity of the trace \eqref{Trace}, 
the gauge transformation law \eqref{eq:WL} 
and the following expression 
for the star commutator of a function $\fhat\zyx{Z}{Y}{x}$ 
with the exponential $e^{i\Mu Z}\,$ 
(see (\ref{eq:f*e(iMZ)}, \ref{eq:e(iMZ)*f}) for details):
\begin{equation}
\fhat\zyx{Z+2\Mu}{Y}{x}\star e^{i\Mu Z}
=
e^{i\Mu Z}\star\fhat\zyx{Z}{Y}{x}
\;.
\end{equation}
The above construction is formal, and hence the well-definiteness of the observables
\eqref{eq:Observable} depends on the properties of the considered 
curve and solution to Vasiliev equations.
The case of the leading order pre-amplitudes will be discussed at the end of this section.

Previously \cite{Colombo:2010fu,Sezgin:2011hq,Colombo:2012jx},
decorated Wilson loops in the \emph{commutative} $\mathcal{X}_4$ space
have been considered within the context of Vasiliev's theory; 
in particular, for trivial loops, these reduce to invariants of
the form 
\begin{equation}
{\mathcal{I}}_{n_0,t}\left(\Mu\right)
:=
{\rm Tr}\,\left[
{\widehat{\Psi}}{}^{\star n_0}  
\star\starpuis{{\hat\kappa}{\hat{\bar\kappa}}}{t}
\star \starexp{i\Mu\widehat{S}}
\right]
\;, 
\end{equation}
for $t$ being zero or one, where we note the insertion of 
the adjoint operator $e_\star^{i\Mu\hat{S}}$.
As these invariants are independent of the choice of
base point in ${\cal X}_4\,$, they have been referred
to as zero-form charges.
Let us show the equivalence between these invariants 
and twisted straight open Wilson lines 
in $\mathcal{Z}_4\,$, 
thereby providing a geometrical underpinning for 
the insertion of the adjoint impurity $e_\star^{i\Mu\hat{S}}$ 
formed out of the deformed oscillator $\hat{S}\,$.
To this end, we consider the straight line
\begin{equation}
L_{2M} : 
[0,1] \rightarrow \mathcal{X}_4\times \mathcal{Z}_4 : 
\sigma \rightarrow (0,2\sigma\Mu^{\underalpha})
\;.
\end{equation}
The open Wilson lines in noncommutative space form an over-complete set of observables for
noncommutative Yang-Mills theory, see  \cite{Gross:2000ba} and references therein.
As stressed in the same reference, the  \emph{straight} open Wilson lines with 
a complete set of adjoint impurities inserted at one end of the Wilson line 
also provide such a set of observables
\footnote{Indeed, considering a general path \eqref{eq:defCurve}, one can  write 
{\setlength{\abovedisplayskip}{0pt}
\setlength{\belowdisplayskip}{0pt}
\begin{equation*}\WLczyx{\mathcal{C}}{Z}{Y}{x}
=
\WLczyx{\mathcal{C}}{Z}{Y}{x}
\star
\left(\WLczyx{L_{2M}}{Z}{Y}{x}\right)_{\star}^{-1}
\star
\WLczyx{L_{2M}}{Z}{Y}{x}
\;,\end{equation*}}
where $\WLczyx{\mathcal{C}}{Z}{Y}{x}
\star
\left(\WLczyx{L_{2M}}{Z}{Y}{x}\right)_{\star}^{-1}$ is an adjoint operator.
}.
In the case of Vasiliev's theory, these observables can be written as follows:
\begin{align}
\label{eq:def WLmxO}
{\widetilde{O}}_{L_{2M}}\left(\Mu\vert x\right)
&=
\int \diffn{4}{Z}\diffn{4}{Y} \hat{O}\zyx{Z}{Y}{x} \star 
\WLczyx{L_{2M}}{Z}{Y}{x}\star\exp(i\Mu Z) 
\\&=
\int \diffn{4}{Z}\diffn{4}{Y} \hat{O}\zyx{Z}{Y}{x} \star 
\starexp{i\Mu\widehat{S}}
\;.
\end{align}
The last equality comes form the following result:
\begin{equation}
\starexp{i\Mu\widehat{S}} = \WLczyx{L_{2M}}{Z}{Y}{x}\star\exp(i\Mu Z)
\;,
\end{equation}
that is proved order by order in powers of the higher spin connection $\widehat{A}$ in Appendix \ref{sec:thmWL}.

From the form of the equations \eqref{EVS1} --- \eqref{EVS5} involving 
$\widehat{S}$ and $\widehat{\Psi}\,$,
one can see that the most general adjoint operator
one can build out of the master fields is equivalent on shell to:
\begin{equation}
\label{eq:gen adj op}
{\hat{O}}_{n_0,t;\,\underalpha(K)} 
:=
\widehat{\Psi}^{\star n_0}
\star\starpuis{{\hat\kappa}{\hat{\bar\kappa}}}{t}
\star\starpuis{\widehat{S}_{\underalpha}}{K}
\;,
\end{equation}
where, as suggested by the indices, the deformed oscillators $\widehat{S}_{\underalpha}$ are symmetrized.
Through \eqref{eq:def WLmxO}, this yields the following form for the most general observable of Vasiliev's theory:
\begin{equation}
\label{eq:def ZFC(S)}
{\mathcal{I}}_{n_0,t;\underalpha(K)}\left(\Mu\right)
=
\int \diffn{4}{Z}\diffn{4}{Y}
\widehat{\Psi}^{\star n_0}
\star\starpuis{{\hat\kappa}{\hat{\bar\kappa}}}{t}
\star\starpuis{\widehat{S}_{\underalpha}}{K}
\star 
\starexp{i\Mu\widehat{S}}
\;.
\end{equation}
However, it turns out that one can obtain all of 
these observables from the evaluation of the 
following ones, viewed as functions of 
$\Mu_{\underline{\alpha}}\,$:
\begin{equation}
\label{eq:def ZFC}
{\mathcal{I}}_{n_0,t}\left(\Mu\right)
=
\int \diffn{4}{Z}\diffn{4}{Y}
\widehat{\Psi}^{\star n_0}
\star\starpuis{{\hat\kappa}{\hat{\bar\kappa}}}{t}
\star 
\starexp{i\Mu\widehat{S}}
\;.
\end{equation}
Indeed, one has
\begin{equation}
\label{eq:gen ZFC(S)}
\left.
(\partial^M_{\underalpha})^K
{\mathcal{I}}_{n_0,t}\left(\Mu\right)
\right\vert_{M=0}
=
\int \diffn{4}{Z}\diffn{4}{Y}
\widehat{\Psi}^{\star n_0}
\star\starpuis{{\hat\kappa}{\hat{\bar\kappa}}}{t}
\star 
(i\widehat{S}_{\underalpha})^{\star K}
\end{equation}
and the observable \eqref{eq:def ZFC(S)}
can be written as an infinite sum of term of 
this form, upon applying (\ref{EVS4},\,\ref{EVS5}) 
repeatedly in order to symmetrise all the 
deformed oscillators.
We will refer to the observables \eqref{eq:def ZFC} 
as \emph{zero-form charges} since they are nothing but 
the observables considered and evaluated
in some special cases in 
\cite{Colombo:2010fu,Sezgin:2011hq,Colombo:2012jx}.
In the weak field expansion scheme,
we can write the leading order 
contribution to the zero-form charges as
\begin{align}
\label{eq:ZFCltem}
{\mathcal{I}}^{(n_0)}_{n_0,t}\left(\Mu\right)
&=
\int \diffn{4}{Z}\diffn{4}{Y}
\starpuis{\Phi\star {\hat\kappa}}{n_0}
\star\starpuis{{\hat\kappa}{\hat{\bar\kappa}}}{t}
\star e^{i\mu z}\star e^{-i\mubar\zbar}
\\&=
\int \diffn{4}{Z}\diffn{4}{Y}
\left(\starprod_{i=1}^{n_0}
\Phi\yyx{(-)^{i+1}y}{\ybar}{x}
\right)
\star e^{ieyz}\star e^{-it\ybar\zbar}
\star e^{i\mu z}\star e^{-i\mubar\zbar}
\;.
\end{align}
The second line was obtained using \eqref{eq:pisfromks} and defining:
\begin{equation}
e:=n_0+t\mod2
\;,\quad 
t\in\{0,1\}
\;,
\end{equation}

Following \cite{Colombo:2012jx}, 
we use the zero-form charges as building blocks 
for \emph{quasi-amplitudes} of various orders. 
To define the quasi-amplitudes of order $n$, 
we expand a given 
zero-form charge to $n$th order in the 
twisted-adjoint weak field $\Phi\,$
and replace, in a way that 
we specify below, the $n$ fields $\Phi\,$ 
with $n$ distinct external twisted-adjoint 
quantities $\Phi_{i}\,$, 
$i=1,\dots,n$, each of which transforming as 
\eqref{eq:GIf tafV} under a diagonal higher 
spin group acting on all $\Phi_i$'s with the 
same parameter.
The quasi-amplitude 
$\mathcal{Q}_{n_0,t}^{(n)}(\Phi_i\vert M)$ 
is now defined unambiguously as the functional of 
$\Phi_1,...,\Phi_{n}$ 
that is totally symmetric in its $n$ arguments and 
obeys:
\begin{equation}
\left.
\mathcal{Q}_{n_0,t}^{(n)}(\Phi_i\vert M)
\right\vert_{\Phi_1=...=\Phi_{n}=\Phi}
=
{\mathcal{I}}^{(n)}_{n_0,t}\left(\Mu\right)
.
\end{equation}
As shown in \cite{Colombo:2012jx,Didenko:2012tv}, 
at the leading order, 
\emph{i.e.} $n=n_0$, these reproduce the 
correlation functions of bilinear operators 
in the free conformal field theory in 
three dimensions for $n=2,3$ and $4\,$.

In this paper, we are interested in more 
fundamental building blocks from the bulk 
point of view, which are referred to as 
\emph{pre-amplitudes} and are defined as follows:
\begin{equation}
\label{eq:PAltemf}
{\mathcal{A}}_{n_0,t}\left(\Phi_{i}\vert\Mu\right)
:=
\int \diffn{4}{Z}\diffn{4}{Y}
\left(\starprod_{i=1}^{n_0}
\Phi_i\yyx{(-)^{i+1}y}{\ybar}{x}
\right)
\star e^{ieyz}\star e^{-it\ybar\zbar}
\star e^{i\mu z}\star e^{-i\mubar\zbar}
\;.
\end{equation}
The prescription to obtain those objects is to replace $\Phi$ 
in the expression \eqref{eq:ZFCltem} of the leading order zero-form charges 
with the different fields $\Phi_{i}$ in a given order,
conventionally with the label growing from left to right.
We will not discuss their generalization to higher orders in perturbation theory.
The aim of this paper is to strengthen the previous results on 
quasi-amplitudes
by extending it to $n_0$-point function
and showing that the correspondence holds already at the level of basic 
cyclic structures.
The relevant cyclic blocks on the CFT side are Wick contractions (see 
Section \ref{sec:CFT}),
while on the bulk side they are precisely the pre-amplitudes \eqref{eq:PAltemf}.

The first step will be to show the invariance of the pre-amplitudes \eqref{eq:PAltemf} under cyclic permutations of the external legs.
To do so, 
we use the decompositions \eqref{eq:k=ky*kz} to split the integrand 
into a $Z$-dependent part $G^t(Z\vert M)$ and a $Y$-dependent part 
to be specified below. From the cyclicity of the trace and the 
mutual star-commutativity of functions of $Y$ and functions of 
$Z\,$, we compute:
\begin{align}
{\mathcal{A}}_{n_0,t}\left(\Phi_{i}\vert\Mu\right)
&=
\int \diffn{4}{Z}\diffn{4}{Y}
\left(\starprod_{i=1}^{n_0}
\pi^{i+1}\Phi_{i}
\right)
\star\starpuis{\kappa_y}{n_0}
\star\starpuis{\kappa_y{\bar\kappa}_y}{t}
\star G^t(Z\vert M)
\nonumber\\&=
\int \diffn{4}{Z}\diffn{4}{Y}
\left(\starprod_{i=2}^{n_0}
\pi^{i+1}\Phi_{i}
\right)
\star\starpuis{\kappa_y}{n_0}
\star\starpuis{\kappa_y{\bar\kappa}_y}{t}
\star G^t(Z\vert M)
\star \Phi_{1}
\nonumber\\&=
\int \diffn{4}{Z}\diffn{4}{Y}
\left(\starprod_{i=2}^{n_0}
\pi^{i+1}\Phi_{i}
\right)
\star\starpuis{\kappa_y}{n_0}
\star\starpuis{\kappa_y{\bar\kappa}_y}{t}
\star \Phi_{1}
\star G^t(Z\vert M)
\nonumber\\&=
\int \diffn{4}{Z}\diffn{4}{Y}
\left(\starprod_{i=1}^{n_0-1}
\pi^{i}\Phi_{i+1}
\right)
\star \left(\pi^{n_0}(\pi\pibar)^{t}\Phi_{1}\right)
\star\starpuis{\kappa_y}{n_0}
\star\starpuis{\kappa_y{\bar\kappa}_y}{t}
\star G^t(Z\vert M)
\nonumber\\&=
\PAltemSA{\Phi_2,...,\Phi_{n_0},(\pi\pibar)^t\Phi_1}
\;,
\end{align}
where the concluding line is obtained by a change of integration
variables given by $y\to-y$. 
Thus, the cyclic invariance follow from the integer-spin projection 
in \eqref{eq:RC tafV}.

We remark that as Weyl- and normal-ordered symbols of 
operators depending either only on $Y$ or only on $Z$ 
are the same, the integrations over $Y$ and $Z$ can
be factorized.
This simplifying scheme was used in \cite{Didenko:2012tv} 
in deriving the leading order quasi-amplitudes for all $n$, 
whereas the earlier results for $n=2,3$ in \cite{Colombo:2012jx}
were based on a different scheme adapted to the weak-field
expansion of Vasiliev's equations to higher order, for which 
there is no obvious factorization of the integrand.
In what follows, we shall follow the latter approach, and 
perform an alternative derivation of the results of \cite{Didenko:2012tv}, 
which thus can be generalized more straightforwardly to computing 
subleading corrections to open Wilson lines in ${\cal Z}_4\,$.

To the latter end, we define $Y$-space momenta 
$\Lambda_{\underalpha}=(\lambda_\alpha,\lambdabar_{\alphadot})$,
that are complex conjugates of each other, 
\emph{i.e.} $(\lambda,\lambdabar)^\dagger = (\lambdabar,\lambda)$,
and which are not affected by $\star$-products.
Assuming that the twisted-adjoint zero-form 
$\Phi\in {\cal S}(\mathcal{Y}_4)\,$, i.e. that it is a rapidly 
decreasing function,
we have the following Fourier transformation relations:
\begin{align}
\label{eq:def fttafZ}
\tilde\Phi(\Lambda)
:&=
\int\frac{\diffn{4}{Y}}{(2\pi)^2}\Phi(Y)\exp{(-i\Lambda Y)}
\;,\\
\Phi(Y)
&=
\int\frac{\diffn{4}{\Lambda}}{(2\pi)^2}\tilde\Phi(\Lambda)\exp{(i\Lambda Y)}
\;.
\end{align}
Upon defining the following mappings:
\begin{equation}
\pi_{\Lambda}(\lambda,\lambdabar) = (-\lambda,\lambdabar)
\;,\quad
\pibar_{\Lambda}(\lambda,\lambdabar) = (\lambda,-\lambdabar)
\;,\quad
\tau_{\Lambda}(\lambda,\lambdabar) = (i\lambda,i\lambdabar)
\;,
\end{equation}
the integer-spin projection and reality condition in \eqref{eq:RC tafV} and 
the minimal bosonic projection \eqref{eq:MBP tafV}, respectively, translate into
\begin{align}
\label{eq:RC fttaf}
\tilde\Phi^\dagger =\pi_\Lambda\tilde\Phi = \pibar_\Lambda\tilde\Phi
\;,\\
\label{eq:MBP fttaf}
\pi_\Lambda\tilde\Phi = \tau_\Lambda\tilde\Phi
\;.
\end{align}
Writing \eqref{eq:PAltemf} in terms of plane waves gives
\begin{equation}
\label{eq:PAfftem}
{\mathcal{A}}_{n_0,t}\left(\Phi_{i}\vert\Mu\right)
=
\int\left(\prod_{j=1}^{{n_0}}\frac{\diffn{4}{\Lambda_j}}{(2\pi)^2}\right) 
\left(\prod_{j=1}^{{n_0}}\tilde\Phi(\Lambda_j)\right)
F_{n_0,t}\left(\Lambda_i\vert M\right)
\;,
\end{equation}
where the quantity $F_{n_0,t}\left(\Lambda_i\vert M\right)$,
which one may think of as a higher spin form factor, is given by 
\begin{equation}
F_{n_0,t}\left(\Lambda_i\vert M\right)
=
\int \diffn{4}{Z}\diffn{4}{Y}
\left(\starprod_{i=1}^{n_0}
e^{i(-1)^{i+1}\lambda_{i} y+i\lambdabar_{i}\ybar}
\right)
\star e^{ieyz}\star e^{-it\ybar\zbar}
\star e^{i\mu z}\star e^{-i\mubar\zbar}
\;.
\end{equation}
In order to perform the star-products appearing above, 
we need some lemmas. 
First of all, star multiplying exponentials of linear expressions in $Y$ and $Z$ one has
\begin{align}
\label{eq:f*e(iMZ)}
\fhat\zyx{Z}{Y}{x}\star e^{i\Mu Z}
&= 
e^{i\Mu Z}\fhat\zyx{Z-\Mu}{Y-\Mu}{x}
\;,\\
\label{eq:e(iMZ)*f}
e^{i\Mu Z}\star \fhat\zyx{Z}{Y}{x}
&= 
e^{i\Mu Z}\fhat\zyx{Z+\Mu}{Y-\Mu}{x}
\;, \\
\fhat\zyx{Z}{Y}{x}\star e^{i\Lambda Y}
&= e^{i\Lambda Y}\fhat\zyx{Z+\Lambda}{Y+\Lambda}{x}
\;,\\
e^{i\Lambda Y}\star \fhat\zyx{Z}{Y}{x}
&= e^{i\Lambda Y}\fhat\zyx{Z+\Lambda}{Y-\Lambda}{x}
\;.
\end{align}
Then one can show recursively that the following equality holds:
\begin{equation}
\starprod_{j=1}^n \exp(i\Lambda_j Y) 
= 
\exp\left(i\sum_{j=1}^{n}\sum_{k=1}^{j-1}\Lambda_k \Lambda_j\right)
\exp\left(i\sum_{j=1}^{n}\Lambda_jY\right)
\;.
\end{equation}
The following relations, valid for any $t$ and $e\,$, 
will be useful as well:
\begin{align}
\label{eq:f*k}
\fhat\zzyyx{z}{-\zbar}{y}{\ybar}{x}\star e^{ieyz}
&=
e^{ieyz}\fhat\zzyyx{(1-e)z-ey}{-\zbar}{(1-e)y-ez}{\ybar}{x}
\;,\\
\label{eq:f*kb}
\fhat\zzyyx{z}{-\zbar}{y}{\ybar}{x}\star e^{-it\ybar\zbar}
&=
e^{-it\ybar\zbar}\fhat\zzyyx{z}{-(1-t)\zbar-t\ybar}{y}{(1-t)\ybar+t\zbar}{x}
\;.
\end{align}
The above relations allow the factorization of $F_{n_0,t}\left(\Lambda_i\vert M\right)$ as follows:
\begin{equation}
F_{n_0,t}\left(\Lambda_i\vert M\right)
=
\PAffq\PAffem\PAfftm
\;,
\end{equation}
where
\begin{align}
\label{eq:PAffq}
\PAffq 
:&= 
\exp\left(i\left[\sum_{i<j}^{n_0} (-1)^{i+j}\lambda_i\lambda_j 
+\sum_{i<j}^{n_0} \lambdabar_i\lambdabar_j\right]\right)
\;,\\
\PAffem 
:&=
\intyz\exp\left[
i(1-e)\left(-\sum_i (-)^i \lambda_i (y-\mu)+\mu z\right)\right]
\nonumber\\
&\qquad \exp\left[ie\left(y-\sum_i (-)^i \lambda_i\right)(z-\mu)
\right]
\nonumber\\
\label{eq:PAffem noYZ}
&=(2\pi)^{4-2e} 
\Big(\delta^2\left(\mu\right)\delta^2(\sum_j(-)^j\lambda_j)\Big)^{1-e}
\;,\\
\PAfftm 
:&=
\intybzb\exp\left(
i(1-t)\left(\sum_i \lambdabar_i (\ybar-\mubar)-\mubar \zbar\right)
+it\left(-\ybar+\sum_i\lambdabar_i\right)(\zbar+\mubar)
\right)
\nonumber\\\label{eq:PAfftm noYZ}&=
(2\pi)^{4-2t}
\Big(\delta^2\left(\mubar\right)\delta^2(\sum_j\lambdabar_j)\Big)^{1-t}
\;.
\end{align}
The above functions have the following behaviour under cyclic 
permutations of the $Y$-space momenta:
\begin{align}
\PAffqSA{\lambda_1,\lambda_2,...,\lambda_{n_0}}{\lambdabar_1,\lambdabar_2,...,\lambdabar_{n_0}}
&=
\PAffqSA{\lambda_2,...,\lambda_{n_0},-(-)^{n_0}\lambda_{1}}{\lambdabar_2,...,\lambdabar_{n_0},-\lambdabar_{1}}
\;,\\
\PAffemSA{\lambda_1,\lambda_2,...,\lambda_{n_0}}
&=
\PAffemSA{\lambda_2,...,\lambda_{n_0},(-)^{n_0}\lambda_{1}}
\;,\\
\PAfftmSA{\lambdabar_1,\lambdabar_2,...,\lambdabar_{n_0}}
&=
\PAfftmSA{\lambdabar_2,...,\lambdabar_{n_0},\lambdabar_{1}}
\;.
\end{align}
Let us point out the fact that $\PAffem$ (resp. $\PAfftm$) is 
given by $(2\pi)^2$ if $e=1$ (resp. $t=1$)), or else it is
given by a delta function that sets, in $\PAffq\,$, the
momentum $\lambda_1$ (resp. $\lambdabar_1$) equal to 
combinations of the other variables 
$\lambda_j\,$, $j=2, \ldots, n_0$. 
As a result, one can write the cyclic property of $F_{n_0,t}\left(\Lambda_i\vert M\right)$ as
\begin{equation}
F_{n_0,t}\left(\Lambda_1,\Lambda_2,...,\Lambda_{n_0}\right)
=
F_{n_0,t}\left(\Lambda_2,...,\Lambda_{n_0},(-)^t\Lambda_1\right)
\;.
\end{equation}
This behaviour under cyclic permutations can be used in \eqref{eq:PAfftem} 
to show the cyclic invariance of ${\mathcal{A}}_{n_0,t}\left(\Phi_{i}\vert\Mu\right)$ as:
\begin{align*}
{\mathcal{A}}_{n_0,t}\left(\Phi_{i}\vert\Mu\right)
:&=
\left(\prod_{n=1}^{{n_0}}\int\frac{\diffn{4}{\Lambda_n}}{(2\pi)^2} \tilde\Phi_n(\Lambda_n)\right)
F_{n_0,t}\left(\Lambda_i\vert M\right)
\\&=
\int\frac{\diffn{4}{\Lambda_1}}{(2\pi)^2}\tilde\Phi_1(\Lambda_{1})
\left(\prod_{n=2}^{n_0}\int\frac{\diffn{4}{\Lambda_n}}{(2\pi)^2} \tilde\Phi_n(\Lambda_n)\right)
F_{n_0,t}\left(\Lambda_2,...,\Lambda_{n_0},(-)^t\Lambda_1\right)
\\&=
\int\frac{\diffn{4}{\Lambda_{n_0}}}{(2\pi)^2}\tilde\Phi_1((-)^{t}\Lambda_{n_0})
\left(\prod_{n=1}^{n_0-1}\int\frac{\diffn{4}{\Lambda_n}}{(2\pi)^2} \tilde\Phi_{n+1}(\Lambda_n)\right)
F_{n_0,t}\left(\Lambda_i\vert M\right)
\\&=
\PAltemSA{\Phi_2,...,\Phi_{n_0},\Phi_1}
\;.
\end{align*}
The third line is a mere change of variables, while the last one 
uses \eqref{eq:RC fttaf}.

The computations of this section have shown that the $M$ dependence of the pre-amplitudes ${\mathcal{A}}_{n_0,t}\left(\Phi_{i}\vert\Mu\right)$ can be factorised as follows:
\begin{equation}
\label{eq:fact M dep}
{\mathcal{A}}_{n_0,t}\left(\Phi_{i}\vert\Mu\right)
=
\delta^2(\mu)^{1-e}\delta^2(\mubar)^{1-t}
{\mathcal{A}}_{n_0,t}\left(\Phi_{i}\right)
\;.
\end{equation}
This result presents the divergences that were first discussed in \cite{Colombo:2012jx}.
This is a consequence of the integrand of \eqref{eq:PAltemf} not being in $L^1(\mathcal{Y}_4\times {\cal Z}_4)$.
Let us replace the twistor plane waves $e^{iMZ}$ by a function $\mathcal{V}(Z)\in \mathcal{S}({\cal Z}_4)$, then the following object is well defined:
\begin{align}
{\mathcal{A}}^{\mathcal{V}}_{n_0,t}\left(\Phi_{i}\right)
:&=
\int \diffn{4}{Z}\diffn{4}{Y}
\left(\starprod_{i=1}^{n_0}
\Phi_i\yyx{(-)^{i+1}y}{\ybar}{x}
\right)
\star e^{ieyz}\star e^{-it\ybar\zbar}
\star \mathcal{V}(Z)
\;,\\
&=
\int \diffn{4}{Z}\diffn{4}{Y}
\mathcal{V}(Z)
\star \left(\starprod_{i=1}^{n_0}
\Phi_i\yyx{(-)^{i+1}y}{\ybar}{x}
\right)
\star e^{ieyz}\star e^{-it\ybar\zbar}
\;,
\end{align}
where the last line follows from the cyclicity of the trace.
Indeed, it can be shown from (\ref{eq:fact trace},\,\ref{eq:f*k},\,\ref{eq:f*kb}) that
\footnote{Let $\fhat$ a function and $A$ and $B$ two sets of functions,
we define
\begin{align*}
A\star B := \left\{\hat{a}\star \hat{b}: \hat{a}\in A, \hat{b}\in B\right\}
\;,\quad
A\star \fhat := \left\{\hat{a}\star \fhat: \hat{a}\in A\right\}
\;.
\end{align*}
}: 
\begin{align}
\mathcal{S}({\cal Z}_4)\star \mathcal{S}(\mathcal{Y}_4)\star e^{ieyz}\star e^{-it\ybar\zbar}
&\subseteq 
L^1({\cal Z}_4)\star L^1(\mathcal{Y}_4)\star e^{ieyz}\star e^{-it\ybar\zbar}
\\&\subseteq 
L^1(\mathcal{Y}_4\times {\cal Z}_4)\star e^{ieyz}\star e^{-it\ybar\zbar}
\\&=
L^1(\mathcal{Y}_4\times {\cal Z}_4)
.
\end{align}
Because of the following analogous of \eqref{eq:def fttafZ}:
\begin{align}
\mathcal{\tilde{V}}(\Mu)
:&=
\int\frac{\diffn{4}{Z}}{(2\pi)^2}\mathcal{V}(Z)\exp{(-iMZ)}
\;,\\
\mathcal{V}(Z)
&=
\int\frac{\diffn{4}{M}}{(2\pi)^2}\mathcal{\tilde{V}}(\Mu)\exp{(iMZ)}
\;.
\end{align}
we have that
\begin{align}
\label{eq:regularisedAmplitude}
{\mathcal{A}}_{n_0,t}^{\mathcal{V}}\left(\Phi_{i}\right)
&= \int\diffn{4}{\Mu}\mathcal{\tilde{V}}(\Mu)
{\mathcal{A}}_{n_0,t}\left(\Phi_{i}\vert\Mu\right)
\\&=
\mathcal{\tilde{V}}_{t,e;\alpha(0),\alphadot(0)}\,{\mathcal{A}}_{n_0,t}\left(\Phi_{i}\right)
\;.
\end{align}
This amounts to the regularisation scheme introduced in \cite{Colombo:2012jx},
where $\mathcal{\tilde{V}}(\Mu)$ was called smearing function.
The introduction of the (field-independant) smearing function does not 
spoil the gauge invariance.
At this order in perturbation theory, its effect on the pre-amplitudes 
${\mathcal{A}}_{n_0,t}^{\mathcal{V}}\left(\Phi_{i}\right)$ is 
the appearance of four coupling constants, a particular case of the 
ones listed in Table \ref{tab:div}.
\begin{table}[h]
\centering
\begin{tabular}{|l|l|l|l|l|}
\hline
$n_0$ &t &e &Divergences 
&$\mathcal{\tilde{V}}_{t,e;\alpha(k),\alphadot(\bar{k})}$
\\\hline
even &1 &1 &None 
&$\int\diffn{2}{\mu}\diffn{2}{\mubar}
(-\mu_\alpha)^{k}
(-\mubar_{\alphadot})^{\bar{k}}
\mathcal{\tilde{V}}
(\mu,\mubar)$
\\
odd &1 &0 &$\delta^2\left(\mu\right)$
&$
\int\diffn{2}{\mubar}
(-\mubar_{\alphadot})^{\bar{k}}
(i\partial_\alpha)^k
\mathcal{\tilde{V}}
(0,\mubar)$
\\
odd &0 &1 &$\delta^2\left(\mubar\right)$
&$
\int\diffn{2}{\mu}
(-\mu_\alpha)^{k}
(i\partial_{\alphadot})^{\bar{k}}
\mathcal{\tilde{V}}
(\mu,0)$
\\
even &0 &0 &$\delta^4\left(\Mu\right)$
&$
(i\partial_\alpha)^k
(i\partial_{\alphadot})^{\bar{k}}
\mathcal{\tilde{V}}
(0,0)$
\\\hline
\end{tabular}
\caption{Summary of the divergent structures in the pre-amplitudes.}
\label{tab:div}
\end{table}

However, this does not mean that those constants are the only contributions of the regularising function that are observable.
Indeed, let us show that the complete information about the function $\mathcal{\tilde{V}}(M)$ 
appear in the evaluation of the pre-amplitudes
${\mathcal{A}}^{\mathcal{V}}_{n_0,t;\alpha(k),\alphadot(\bar{k})}\left(\Phi_{i}\right)$
obtained by applying the  prescription explained below \eqref{eq:PAltemf}
to the most general observables \eqref{eq:def ZFC(S)},
then regularising as in \eqref{eq:regularisedAmplitude}.
Let us stress that even though those 
observables can be built from $
{\mathcal{A}}_{n_0,t}\left(\Phi_{i}\vert\Mu\right)$ 
using \eqref{eq:gen ZFC(S)},
this expression \eqref{eq:gen ZFC(S)} needs 
the $M$ dependent 
(hence divergent) version of the pre-amplitudes
${\mathcal{A}}_{n_0,t}\left(\Phi_{i}\vert\Mu\right)$, 
and does not apply with the regularised one.
This being said, the non-regularised version of ${\mathcal{A}}^{\mathcal{V}}_{n_0,t;\alpha(k),\alphadot(\bar{k})}\left(\Phi_{i}\right)$ reads:
\begin{align}
{\mathcal{A}}_{n_0,t;\underalpha(K)}\left(\Phi_i\vert\Mu\right)
&=
\int \diffn{4}{Z}\diffn{4}{Y}
\left(\starprod_{i=1}^{n_0}\hat{\Psi_i}\right)
\star\starpuis{{\hat\kappa}{\hat{\bar\kappa}}}{t}
\star (Z_{\underalpha})^K
\star e^{iMZ}
\nonumber\\&=
\int \diffn{4}{Z}\diffn{4}{Y}
\left(\starprod_{i=1}^{n_0}\hat{\Psi_i}\right)
\star\starpuis{{\hat\kappa}{\hat{\bar\kappa}}}{t}
\star (Z_{\underalpha}-M_{\underalpha})^K e^{iMZ}
\nonumber\\&=
\int \diffn{4}{Z}\diffn{4}{Y}
\left(\starprod_{i=1}^{n_0}\hat{\Psi_i}\right)
\star\starpuis{{\hat\kappa}{\hat{\bar\kappa}}}{t}
\star(-i\partial^M_{\underalpha}-M_{\underalpha})^K e^{iMZ}
\nonumber\\&=
(-i\partial^M_{\underalpha}-M_{\underalpha})^K
{\mathcal{A}}_{n_0,t}\left(\Phi_{i}\vert\Mu\right)
\;.
\end{align}
From \eqref{eq:fact M dep}, we can extract the contributions 
where $(Z_{\underline{\alpha}})^{K}$ 
brings 
$z_{\alpha}^k$ $\bar z_{\dot \alpha}^{\bar k}\,$: 
\begin{equation}
{\mathcal{A}}_{n_0,t;\alpha(k),\alphadot(\bar{k})}\left(\Phi_{i}\vert\Mu\right)
=
\left((-i\partial_\alpha)^k\delta^2(\mu)\right)^{1-e}
(-\mu_\alpha)^{ke}
\left((-i\partial_{\alphadot})^{\bar{k}}\delta^2(\mubar)\right)^{1-t}
(-\mubar_{\alphadot})^{\bar{k}t}
{\mathcal{A}}_{n_0,t}\left(\Phi_{i}\right)
\;.
\end{equation}
Let us make precise that the absence of powers of $\mu$ in the $e=1$ case 
is due to the symmetrization of the indices,
yielding the commutation of $\mu$ and $\partial_\mu$,
which in turn allows to put all $\mu$ right in front of the delta function.
The same argument rules out the presence of $\mubar$ when $t=1\,$, $\partial_\mu$ when $e=0$ and $\partial_{\mubar}$ when $e=1$.
After various integrations by part, the regularised general pre-amplitudes read:
\begin{equation}
{\mathcal{A}}^{\mathcal{V}}_{n_0,t;\alpha(k),\alphadot(\bar{k})}\left(\Phi_{i}\right)
=
\mathcal{\tilde{V}}_{t,e;\alpha(k),\alphadot(\bar{k})}
{\mathcal{A}}_{n_0,t}\left(\Phi_{i}\right)
\;,
\end{equation}
where, again, the prefactors are listed in Table \ref{tab:div}.
Once again, the well-definiteness of each of those observables relies on the properties of $\mathcal{V}\,$.
All of them are finite since $\mathcal{V}$ is a rapidly decreasing function.

In the next section, we will be less general and see what happens when plugging a precise weak-field in the definition of the pre-amplitude.
In that context, the question of the $M$ dependence will be irrelevant
and we will only be interested in computing 
${\mathcal{A}}_{n_0,t}\left(\Phi_{i}\right)$ from the following equivalent of \eqref{eq:PAfftem}:
\begin{equation}
\label{eq:PAffte}
{\mathcal{A}}_{n_0,t}\left(\Phi_{i}\right)
=
\int\left(\prod_{j=1}^{{n_0}}\frac{\diffn{4}{\Lambda_j}}{(2\pi)^2}\right) 
\left(\prod_{j=1}^{{n_0}}\tilde\Phi(\Lambda_j)\right)
F_{n_0,t}\left(\Lambda_i\right)
\;,
\end{equation}
where $F_{n_0,t}\left(\Lambda_i\right)$ is defined by
\begin{equation}
\label{eq:factor F(M)}
F_{n_0,t}\left(\Lambda_i\vert M\right)
=
\delta^2(\mu)^{1-e}\delta^2(\mubar)^{1-t}
F_{n_0,t}\left(\Lambda_i\right)
\;.
\end{equation}

\section{Correlators from zero-form charges}
\label{sec:Correlators}

The purpose of this section is to use the 
bulk-to-boundary propagators of \cite{Giombi:2009wh,Giombi:2010vg} as weak fields
inside the expression \eqref{eq:PAffte} for the pre-amplitudes, 
and fully evaluate the complete expression that results.

Then, in Section \ref{sec:CFT} we will separately compute the 
$n$-point correlation functions of conserved currents of the
free $CFT_3$ corresponding to a set of free bosons.
The latter model was conjectured in \cite{Sezgin:2002rt} 
to be dual to the type-A Vasiliev model (with a parity-even bulk 
scalar field) where the bulk scalar field obeys the Neumann 
boundary condition, sometimes called ``irregular boundary condition''. 
We will show that both expressions, i.e. the pre-amplitude for Vasiliev's 
equations on the one hand  and the correlations functions
of conserved currents in the free $CFT_3$ on the other hand, 
exactly coincide.
We stress that the pre-amplitude \eqref{eq:PAffte}
only refers to the free Vasiliev equations and does not take 
into account the interactions incorporated into the fully 
nonlinear model.

After \cite{Sezgin:2002rt}, Klebanov and 
Polyakov \cite{Klebanov:2002ja}
further conjectured that the Vasiliev model where the 
scalar field is parity-even and obeys Dirichlet boundary condition 
should be dual to the critical $U(N)$ model. 
As for the free $U(N)$ model, we instead stick to the Neumann
boundary condition for the bulk scalar field of the Vasiliev model, 
and use the boundary to bulk propagators with the conventions 
of \cite{Colombo:2012jx} that we are now going to recall.

The metric of $AdS$ is expressed in Poincar\'e coordinates, 
so that $\text{d}s^2 = \frac{1}{r^2}\eta_{\mu\nu}\dx^\mu\dx^\nu\,$.
One of the space-like coordinates in $x^\mu$ is the radial 
variable $r$ that vanishes on the boundary of $AdS\,$.
The Minkowski metric components 
$\eta_{\mu\nu}$ and the inverse $\eta^{\mu\nu}$ are used to lower 
and raise world indices. We do not use the components of 
the complete metric $\text{d}s^2\,$, as one might expect
in a geometric formulation. 
We caracterise vectors that are tangent to the boundary 
by the vanishing of their $r$-component.
In the same spirit, all spinors have a bulk notation, 
with dotted and undotted indices, and boundary Dirac spinors 
will be defined as the boundary value 
of bulk spinors submitted to an appropriate projection that 
we specify below.
We use the four matrices $\sigma^\mu\,$, three of which 
are the Pauli matrices, to link any vector 
$v^\mu$ to a $2\times2$ Hermitian matrix as 
$v_{\alpha\betadot}=\bar{v}_{\betadot\alpha}=v_\mu(\sigma^\mu)_{\alpha\betadot}\,$. 
As before, we will omit most of the spinorial indices and do all 
contractions according to the NW-SE convention.

To every pair of points of $AdS_4$ with respective coordinates $x_i^\mu$ 
and $x_j^\mu\,$, we can associate the following two sets 
of quantities:
\begin{align}
\GYsv{i}{j}^{\mu} = x_i^\mu - x_j^\mu
\qquad {\rm and} \qquad 
\GYisv{i}{j}^{\mu} = (\GYsv{i}{j})^{-2}\GYsv{i}{j}^{\mu}
\;.
\end{align}
From now on, all the points will be taken on the boundary, 
except for one bulk point with coordinates $x^\mu_0$ 
(in particular, its $r$ coordinate will be denoted $r_0$).
Of interest in this section will be the following $2\times2$
matrices
\begin{equation}
\label{eq:def CS}
\CSu{i}:=\sigma^{r}-2r_0\GYoisv{i}
\;, 
\end{equation}
that are attached to every boundary point with coordinate $x_i\,$. 
Some of the properties of the matrices $\CSu{i}$ are collected 
in Appendix \ref{app:CS} and will implicitly  be used in the rest 
of this section. Among them is:
\begin{equation}
\CSdm{i}{j}
:=
\det\left(\CSu{i}-\CSu{j}\right)
=
\frac{4r_0^2(\GYsv{i}{j})^2}{(\GYosv{i})^2(\GYosv{j})^2}
\;.
\end{equation}

The propagator of the spin-$s$ component of the master field $\Phi$ from 
the chosen bulk point to a given boundary point of coordinates $x_i^\mu$ 
was computed in \cite{Giombi:2009wh}.
The boundary conditions were chosen as follows: Neumann for the scalar 
field and Dirichlet for the spin-$s>0$ fields.
For the case of the bosonic model, the bulk-to-boundary 
propagator of the master field $\Phi$ was given 
in \cite{Colombo:2012jx} by:
\begin{equation}
\GYbtbY{i}(x_0,x_i,\GYpsu{i}\vert Y)
:=
\GYbtbpf{i}e^{iy\CSu{i}\ybar}
\sum_{\GYsnO{i}=\pm1}\left(
e^{i\theta} e^{i\GYsnO{i}\GYpnb{i}\CSb{i}y}
+e^{-i\theta} e^{i\GYsnO{i}\GYpnu{i}\CSu{i}\ybar}
\right)
\;,
\end{equation}
where $\GYpsu{i}$ denotes the polarization spinor attached 
to the boundary point $x_i\,$, and where
\begin{equation}
\label{eq:def K,khi,nu}
\GYbtbpf{i} := (\GYosv{i})^{-2}r_0
\;,\quad
\GYpnu{i}:=\sqrt{2r_0}\,\CSu{i}\,\GYoisvb{i}\,\GYpsu{i}
\;,\quad
(\GYpsu{i})^{\dagger} = \GYpsb{i} = {\bar\sigma}^{r}\GYpsu{i}
\;,\quad
(\GYpnu{i})^{\dagger} = \GYpnb{i} = -\CSb{i}\GYpnu{i}
\;.
\end{equation}
The propagator $\GYbtbY{i}(x_0,x_i,\GYpsu{i}\vert Y)$
is an imaginary Gaussian in ${\cal Y}_4\,$
Hence, submitted to the usual $i\epsilon$ prescription
allowing the use of \eqref{eq:wiki gaussian},
it becomes a rapidly decreasing function,
justifying the above procedure.
Following the definition \eqref{eq:def fttafZ}, 
the Fourier transform of 
$\GYbtbY{i}(x_0,x_i,\GYpsu{i}\vert Y)$ is given by
\begin{align}
\GYbtbL{j}(x_0,x_j,\GYpsu{j}\vert \Lambda_j)
&=
\GYbtbpf{j}e^{i\lambda_j\CSu{j}\lambdabar_j}
\sum_{\GYsnO{j}=\pm1}\left(
e^{-i\theta} e^{i\GYsnO{j}\GYpnu{j}\lambda_j}
+e^{i\theta} e^{i\GYsnO{j}\GYpnb{j}\lambdabar_j}
\right)
\\\label{eq:GYbtbL}&=
\GYbtbpf{j}e^{i\lambda_j\CSu{j}\lambdabar_j}
\sum_{\GYsau{j}\in\{0,1\}}
\sum_{\GYsnO{j}=\pm1}
e^{i\theta\left(1-2\GYsau{j}\right)} 
e^{i\GYsau{j}\GYsnO{j}\GYpnu{j}\lambda_j
+ i(1-\GYsau{j})\GYsnO{j}\GYpnb{j}\lambdabar_j}
\;.
\end{align}
Let us emphasize that $\GYbtbL{i}$ satisfies the reality conditions \eqref{eq:RC fttaf} but not the minimal bosonic projection. 
We will take care of this projection separately at the end of this section.

Now we insert this propagator into \eqref{eq:PAffte}, which yields:
\begin{equation}
\label{eq:PAltemP1}
{\mathcal{A}}_{n_0,t}\left(\GYbtbY{i}\right)
=
\int\left(\prod_{j=1}^{{n_0}}\int\frac{\diffn{4}{\Lambda_j}}{(2\pi)^2}\right) \left(\prod_{j=1}^{n_0}\GYbtbL{j}(\Lambda_j)\right)
F_{n_0,t}\left(\Lambda_i\right)
\;.
\end{equation}

\vspace{.3cm}
Let us introduce the following notation :
\begin{equation}
\label{eq:sum s,e}
\sum_{\sigma,\varepsilon}:=
\sum_{\GYsau{1}\in\{0,1\}} 
\sum_{\GYsnO{1}=\pm1}
...
\sum_{\GYsau{n_0}\in\{0,1\}} 
\sum_{\GYsnO{n_0}=\pm1}
\end{equation}
By making use of 
(\ref{eq:PAffq}, \ref{eq:PAffem noYZ}, \ref{eq:PAfftm noYZ},\ref{eq:factor F(M)}), 
we can write the 
expression \eqref{eq:PAltemP1} in the following way:
\begin{align}
&{\mathcal{A}}_{n_0,t}\left(\GYbtbY{i}\right) =  
\label{4.9}\\&
\alpha_{n_0,t}\,\sum_{\sigma,\varepsilon}
e^{-i\theta\lilsum_i(2\GYsau{i}-1)} 
\int\diffn{4n_0}{\Lambda}
e^{\frac{i}{2}\transp\Lambda R'\Lambda+i\transp{J'}\Lambda}
\;
\Big[\delta^2(\sum_{j=1}^{n_0}(-)^j\lambda_j)\Big]^{(1-e)}\,
\Big[ \delta^2(\sum_{j=1}^{n_0}\lambdabar_j)\Big]^{(1-t)} \;,
\nonumber 
\end{align}
where the pre-factor in the above expression is given by
\begin{equation}
\label{eq:Gpftem}
\alpha_{n_0,t} :=
(2\pi)^{8-2n_0-2e-2t}
\left(\prod_{i=1}^{n_0}\GYbtbpf{i}\right)
\;,
\end{equation}
and the symbol $\Lambda=(\lambda_i,\lambdabar_{\ibar})$,
where $i$ and $\ibar$ run from $1$ to $n_0$,
denotes a $4 n_0$-dimensional column vector\footnote{We recall that each $\lambda_i$ and $\bar\lambda_{\ibar}$ carries a Weyl spinor index.}.
Then the entries of the matrix $R'$ and the source 
$J'=({j}{\,}'{\!}_i , \bar{\jmath}{\,}'{\!}_{\bar \imath})$ 
are given by
\begin{align}
R'_{ij} =& (1-\Bk{i}{j})(-)^{i+j+\Theta(i,\,j)}
\;,\quad
R'_{\ibar\jbar} = (1-\Bk{\ibar}{\jbar})(-)^{\Theta(\ibar,\,\jbar)}
\;,\\
R'_{i\jbar} =& \Bk{i}{\jbar}\CSu{i}
\;,\quad
R'_{\ibar j} = \Bk{\ibar}{j}\CSb{\ibar}
\;,\quad
{j}{\,}'{\!}_{i} = \GYsau{i}\GYsnO{i}\GYpnu{i}
\;,\quad
\bar{\jmath}{\,}'{\!}_{\ibar}= (1-\GYsau{\ibar})\GYsnO{\ibar}\GYpnb{\ibar}
\;.
\end{align}
In this expression, $\Theta(x,\,y)$ is a function 
whose value is 1 when $x$ is greater than $y$
and $0$ otherwise.
Since it always comes in expressions multiplied by $(1-\delta_{xy})$,
we do not need to specify the value of $\Theta(x,\,x)$.
In the case when $e=0$ (resp. $t=0$), in \eqref{4.9} 
we integrate out $\lambda_{n_0}$ (resp. $\lambdabar_{n_0}$) 
so that ${\mathcal{A}}_{n_0,t}\left(\GYbtbY{i}\right)$ is given by
\begin{equation}
{\mathcal{A}}_{n_0,t}\left(\GYbtbY{i}\right)=
\alpha_{n_0,t}
\sum_{\sigma,\varepsilon}
e^{-i\theta\lilsum_{j=1}^{n_0}(2\GYsau{j}-1)}
\int\diffn{2n+2\bar{n}}{\Lambda}
e^{\frac{i}{2}\transp\Lambda R\,\Lambda+i\transp{J}\,\Lambda}
\;,
\label{4.13}
\end{equation}
where the matrix $R$ and the source $J$ are given by
\begin{align}
R_{ij} =&\; (1-\Bk{i}{j})(-)^{i+j+\Theta(i,\,j)}
\;,\quad
R_{\ibar\jbar} = (1-\Bk{\ibar}{\jbar})(-)^{\Theta(\ibar,\,\jbar)}
\;,\\
R_{i\jbar} =&\; \Bk{i}{\jbar}\CSu{i}
+\left(
-(1-t)\delta_{i,n_0}+(1-e)\delta_{\jbar,n_0}(-)^{i}
+(1-t)(1-e)(-)^i\right)\CSu{n_0}
\;,\\
R_{\ibar j} =&\; \Bk{\ibar}{j}\CSb{\ibar}
+\left(
-(1-t)\delta_{j,n_0}+(1-e)\delta_{\ibar,n_0}(-)^{j}
+(1-t)(1-e)(-)^j\right)\CSb{n_0}
\;,\\\label{eq:Gjad}
{j}\,_{i} =&\; \GYsau{i}\GYsnO{i}\GYpnu{i} - 
(1-e)(-)^{i+n_0}\GYsau{n_0}\GYsnO{n_0}\GYpnu{n_0}
\;,\quad
\bar{\jmath}\,_{\ibar} = (1-\GYsau{\ibar})\,\GYsnO{\ibar}\,\GYpnb{\ibar} - 
(1-t)(1-\GYsau{n_0})\,\GYsnO{n_0}\,\GYpnu{n_0}
\;,
\end{align}
where the indices $i$ and $\ibar$ 
labeling the entries of the matrix $R$ and the column vector $J$ 
now run over the following values :
\begin{equation}
i\in\{1,...,n=n_0-(1-e)\}
\;,\quad
\ibar\in\{1,...,\nbar=n_0-(1-t)\}
\;.
\end{equation}
The Gaussian integration in \eqref{4.13}
can be carried out\footnote{It is possible to make a transformation
$\Lambda \rightarrow \Omega \Lambda$ such that 
$\tilde{R} := \Omega^T R \Omega$ is a \emph{real} symmetric 
matrix and therefore can be diagonalised by an orthogonal matrix.} 
via the formula 
\begin{equation}
\label{eq:wiki gaussian}
\mathcal{G} :=
\int\diffn{2n+2\bar{n}}{\Lambda}
e^{\frac{i}{2}\transp\Lambda R\Lambda+i\transp{J}\Lambda}
=
\sqrt{\frac{(2i\pi)^{2n+2\bar{n}}}{\det R}}
e^{\frac{i}{2}\transp{J}R^{-1}J}
\;,
\end{equation}
which differs from the usual gaussian integration formula by a change 
of sign due to the NW-SE convention.
As we show in Appendix \ref{app:gauss}, 
the determinant and inverse of $R$ are given by 
\begin{align}
\label{eq:Gdmad}
\det R &= 
2^{4(n_0-1)}r_0^{2n_0}\prod_{i=1}^{n_0}
\frac{(\GYsv{i}{i+1})^2}{(\GYosv{i})^4}
\;,\\
R^{-1}_{ij} &=
\sum_{\eta=\pm1}\frac{1}{2\CSdm{i}{i+\eta}}
\left(-\eta\Bk{i}{j}+\Bk{i+\eta}{j}\xi_{i,i+\eta}\right)
(\CSu{i}-\CSu{i+\eta})\CSb{i+\eta}
\;,\\
R^{-1}_{\ibar\jbar} &=
\sum_{\eta=\pm1}\frac{1}{2\CSdm{\ibar}{\ibar+\eta}}
\left(-\eta\Bk{\ibar}{\jbar}-\Bk{\ibar+\eta}{\jbar}\xi_{\ibar,\ibar+\eta}\right)
(\CSb{\ibar}-\CSb{\ibar+\eta})\CSu{\ibar+\eta}
\;,\\
R^{-1}_{i\jbar} &=
\sum_{\eta=\pm1}\frac{1}{2\CSdm{i}{i+\eta}}
\left(\Bk{i}{\jbar}+\eta\,\Bk{i+\eta}{\jbar}\,\xi_{i,i+\eta}\right)
(\CSu{i}-\CSu{i+\eta})
\;,\\
\label{eq:Gimad last}
R^{-1}_{\ibar j} &=
\sum_{\eta=\pm1}\frac{1}{2\CSdm{\ibar}{\ibar+\eta}}
\left(\Bk{\ibar}{j}-\eta\,\Bk{\ibar+\eta}{j}\,\xi_{\ibar,\ibar+\eta}\right)
(\CSb{\ibar}-\CSb{\ibar+\eta})
\;,
\end{align}
where it should be understood that the indices $j$ and $j+kn_0$ are identified with each other for any integers $j$ and $k$,
and where the coefficients $\xi_{i,i+\eta}$ are defined as follows:
\begin{align}
\xi_{i,i+\eta}
&=
-\eta + t\;\delta_{i,n_0}\,(\eta+1)+t\;\delta_{i+\eta,n_0}\,(\eta-1)
\;.
\end{align}
Using the above expressions for the inverse matrix $R^{-1}$ and 
the source $J$ into \eqref{eq:wiki gaussian}, we get
\begin{align}
\label{eq:Gint}
\mathcal{G} 
&= 
\sqrt{\frac{(2i\pi)^{2n+2\bar{n}}}{2^{4(n_0-1)}r_0^{2n_0}}
\prod_{i=1}^{n_0}\frac{(\GYosv{i})^4}{(\GYsv{i}{i+1})^2}}
\;\exp\left(
-\frac{i}{4}\sum_{i=1}^{n_0}\GYciQ{i}{i-1}{i+1}\right)
\mathcal{G}_P
\;,\\\label{eq:GintP}
\mathcal{G}_P &=
\exp\left(
-\frac{i}{2}\sum_{i=1}^{n_0}
(-)^{t\,\delta_{i,n_0}}
\GYsnO{i}\GYsnO{i+1}
\left(2\GYsau{i+1}-1\right)
\GYciP{i}{i+1}
\right)
\;.
\end{align}
Where the conformally-invariant variables are defined as in \cite{Colombo:2012jx} :
\begin{equation}
\label{eq:def PQ bulk}
\GYciP{i}{i+1}= \GYpsu{i}\,\sigma^r\,\GYisvb{i}{i+1}\,\GYpsu{i+1}
\;,\quad 
\GYciQ{i}{i-1}{i+1} = \GYpsu{i}\,\sigma^{r}\,
\left(\GYisvb{i}{i+1}-\GYisvb{i}{i-1}\right)\,\GYpsu{i}
\;.
\end{equation}
At this stage, as can be seen in \eqref{4.13},
the next step is to sum over all values taken by 
$\sigma_1,\ldots,\sigma_{n_0}\,$.
In order to do so,
one can show the following two 
identities, holding for any~$\GYciPt{i}{j}\,$:
\begin{align}
\sum_{\GYsnO{2},...,\GYsnO{(n_0-1)}}
e^{\frac{i}{2}\sum_{i=1}^{n_0-1}\GYsnO{i}\GYsnO{i+1}\GYciPt{i}{i+1}}
&\equiv
2^{n_0-2}\left(
\prod_{i=1}^{n_0-1}\cos\left(\tfrac{1}{2}\GYciPt{i}{i+1}\right)
+i^{n_0-1}\GYsnO{1}\GYsnO{n_0}\prod_{i=1}^{n_0-1}\sin\left(\tfrac{1}{2}\GYciPt{i}{i+1}\right)
\right)
\;,\\
\sum_{\GYsnO{1},...,\GYsnO{n_0}}
e^{\frac{i}{2}\sum_{i=1}^{n_0}\GYsnO{i}\GYsnO{i+1}\GYciPt{i}{i+1}}
&\equiv
2^{n_0}\left(
\prod_{i=1}^{n_0}\cos\left(\tfrac{1}{2}\GYciPt{i}{i+1}\right)
+i^{n_0}\prod_{i=1}^{n_0}\sin\left(\tfrac{1}{2}\GYciPt{i}{i+1}\right)
\right)
\;,
\end{align}
where the first identity can be derived recursively on $n_0$
and the second one can be obtained from the first relation  
by summing over $\sigma_1$ and $\sigma_{n_0}\,$.
Replacing $\GYciPt{i}{i+1}$ by 
$(-1)^{t\delta_{i,n_0}}\left(2\GYsau{i+1}-1\right)\GYciP{i}{j}\,$,
one finds
\begin{equation}
\sum_{\sigma_1,\ldots,\sigma_{n_0}} \mathcal{G}_P
= 
2^{n_0}
\prod_{i=1}^{n_0}\cos\left(\tfrac{1}{2}\GYciP{i}{i+1}\right)
-(2i)^{n_0}(-1)^{t}
\left(\prod_{j=1}^{n_0}\left(1-2\GYsau{j}\right)\right)
\prod_{i=1}^{n_0}\sin\left(\tfrac{1}{2}\GYciP{i}{i+1}\right)
\;.
\end{equation}
At this stage, all we have to do is to sum over the 2 values 
(zero and one) taken by each of the variables 
$\varepsilon_i\,$, $i=1,\ldots,n_0\,$,
or equivalently summing over the two values $\pm1$ taken by 
the $n_0$ variables $(1-2\epsilon_i)\,$, $i=1,\ldots,n_0\,$, 
so as to yield 
\begin{align}
\sum_{\sigma,\varepsilon}
e^{i\theta\lilsum_i(1-2\GYsau{i})}
\mathcal{G}_P
\label{eq:Gssb GintP}&=
2^{2n_0}
\left(
(\cos\theta)^{n_0}
\prod_{i=1}^{n_0}\cos\left(\tfrac{1}{2}\GYciP{i}{i+1}\right)
-(-1)^{t}(\sin\theta)^{n_0}
\prod_{i=1}^{n_0}\sin\left(\tfrac{1}{2}\GYciP{i}{i+1}\right)
\right)
\;.
\end{align}

Gathering all the prefactors appearing in 
\eqref{eq:Gpftem}, \eqref{eq:Gint} and \eqref{eq:Gssb GintP}, 
we finally obtain the following expression for the 
pre-amplitudes ${\mathcal{A}}_{n_0,t}\left(\GYbtbY{i}\right)$
\begin{align}
\label{eq:result PAb}
{\mathcal{A}}_{n_0,t}\left(\GYbtbY{i}\right)
&=
\beta_{n_0,t}\,
\exp\left(
-\frac{i}{4}\sum_{i=1}^{n_0}\GYciQ{i}{i-1}{i+1}\right)
\left(
\prod_{i=1}^{n_0}\frac{1}{\left\vert\GYsv{i}{i+1}\right\vert}\right)
\nonumber\\&\quad\times
\left(
(\cos\theta)^{n_0}
\prod_{i=1}^{n_0}\cos\left(\tfrac{1}{2}\GYciP{i}{i+1}\right)
-(-1)^{t}(\sin\theta)^{n_0}
\prod_{i=1}^{n_0}\sin\left(\tfrac{1}{2}\GYciP{i}{i+1}\right)
\right)
\;,\\
\beta_{n_0,t}
:&=
4(i)^{2n_0-2+e+t}(2\pi)^{2+e+t}
\prod_{j=1}^{n_0}{\rm sgn}(\GYsv{j}{j+1}^2)
\;,
\end{align}
where ${\rm sgn}(x)$ is the sign function.
This expression is one of the central results of the paper. 
It reproduces, by restricting to the case where $t=0\,$ and 
up to constant coefficients, the expression 
obtained by combining the equations (6.19) and (6.20)
of \cite{Didenko:2012tv}. We generalise this result to 
the cases where the pre-amplitudes have extra insertions 
of $(\hat\kappa\hat{\bar\kappa})\,$, see \eqref{eq:ZFCltem}, 
which corresponds to taking $t=1\,$. 
However, we see that at the leading order the extra insertion 
has no effect on the final result, except for a global sign in 
the B-model.
The dependence on $\theta$ was kept as a matter of convenience 
during the computation, but the result should be understood
to hold only for the parity-invariant cases, i.e. for 
$\theta = 0$ (type A model) and $\theta = \pi/2$ 
(type B model). It would be interesting to understand 
how to modify the twisted open Wilson line in order to 
capture genuinely parity-breaking terms. 

\vspace{.3cm}
If one wants to restrict to the minimal bosonic model,
one has to use bulk to boundary propagators $\GYbtbLmb{i}$ satisfying the 
minimal bosonic projection \eqref{eq:MBP fttaf}.
Since it is not the case of the one defined in \eqref{eq:GYbtbL}, 
we have to project it explicitly by defining the propagator for the minimal model as
\begin{align}
\GYbtbLmb{i}(x_0,x_i,\GYpsu{i}\vert \Lambda_i)
:&=
\tfrac{1}{2}\sum_{\xi=0,1}
(\pi_{\Lambda}\tau_{\Lambda})^{\xi}\GYbtbL{i}(x_0,x_i,\GYpsu{i}\vert \Lambda_i)
\\&=
\label{eq:MB prop}
\tfrac{1}{2}\sum_{\xi=0,1}
\GYbtbL{i}(x_0,x_i,i^{\xi}\GYpsu{i}\vert \Lambda_i)
\\&=:
\tfrac{1}{2}\sum_{\xi=0,1}
\tau_{\GYpsu{i}}^{\xi}
\GYbtbL{i}(x_0,x_i,\GYpsu{i}\vert \Lambda_i)
\;.
\end{align}
Then, the pre-amplitude for the minimal bosonic model is given in terms of the non-minimal one as
\begin{equation}
{\mathcal{A}}^{MB}_{n_0,t}\left(\GYbtbY{i}\right)
=
\left(
\prod_{i=1}^{n_0}
\tfrac{1}{2}\sum_{\xi_i=0,1}
\tau_{\GYpsu{i}}^{\xi_i}
\right)
{\mathcal{A}}_{n_0,t}\left(\GYbtbY{i}\right)
\;.
\end{equation}
We will now compute the correlation functions of 
conserved currents on the free $CFT_3$ side, and show that, 
before performing Bose-symmetrisation, the result 
(see \eqref{eq:CFTpa4} below) exactly 
reproduces the formula \eqref{eq:result PAb}. 

\section{Free $U(N)$ and $O(N)$ vector models}
\label{sec:CFT}

The purpose of this section is to compute cyclic building 
blocks for the amplitudes in the free U(N) vector model in a 
space-time of dimension 
$d > 2\,$,
thereby proving explicitly a formula conjectured in \cite{Sleight:2016dba}, where the 3-point functions where computed.
In the three-dimensional case,
we find that they match the pre-amplitudes \eqref{eq:result PAb} 
defined in Vasiliev's bosonic type-A model.
Eventually we will show that the minimal bosonic projection of Vasiliev's type-A model amounts to considering the $O(N)$ vector model.

As just stated, the computations of this section take place in a $d$-dimensional spacetime, whose world indices we will denote 
by Greek letters. This should not create any confusion with 
the other sections where we also use Greek letters but for base 
indices in $d+1$ dimensions.
Let 2 $d$-dimensional vectors $a = a^\mu \frac{\partial}{\partial x^{\mu}}$ and 
$b = b^\mu \frac{\partial}{\partial x^{\mu}}\,$. 
In this section we will use the notation $a\cdot b:= a^\mu b_\mu$ and $a^2:=a^\mu a_\mu\,$.

The fields of the theory are complex Lorentz scalars 
$\phi^{i}$ carrying an internal index $i$.
The theory is free and the propagators are
given by
\begin{equation}
\CFTwc{\phi^{i}(x)}{\phi^{j}(y)}
= 0 = 
\CFTwc{\cc{\phi}_{i}(x)}{\cc{\phi}_{j}(y)}
\;,\quad 
\CFTwc{\phi^{i}(x)}{\cc{\phi}_{j}(y)} =
c_1\,\frac{\delta^i_{\phantom{i}j}}{\left\vert x-y\right\vert^{d-2}}\,
\;,
\end{equation}
where $\left\vert x\right\vert:=\sqrt{x^2}\,$.
This theory is known to be conformal.
The conserved current of spin $s$ is a traceless tensor $J_{\mu(s)}$ containing $s$ derivatives and a single trace in the sense of the internal algebra.
Using a polarisation vector $\epsilon^{\mu}$ and some weights 
$a_s$, one can gather all the conserved currents into a 
generating function, see e.g. 
\cite{Craigie:1983fb,Giombi:2009wh}, 
\cite{Giombi:2016ejx} and references therein: 
\begin{equation}
\label{eq:def J}
\sum_{s=0}^{\infty} a_{s}J_{\mu(s)}(x)\left(\epsilon^{\mu}\right)^{s} 
=
J(x,\epsilon) 
=
\cc{\phi}_i(x)f\left(\epsilon,
\overleftarrow{\partial},
\overrightarrow{\partial}
\right)\phi^i(x)
\;.
\end{equation}
We assume that the function $f$ is analytical.
This section will involve sums over integer values, 
that will always be taken from zero to infinity upon 
identifying the inverse of diverging factorials with zero.
Since the generating function $J(x,\epsilon)$ is Lorentz invariant and the spin $s$ current $J_{\mu(s)}$ contains $s$ derivatives, 
the function $f(\epsilon,u,v)$ can be written as
\begin{equation}
\label{eq:series f}
f(u,v,\epsilon)
=
\sum_{k,\ell,m,p,q}f_{k,\ell,m,p,q}
(\epsilon\cdot u)^k(\epsilon\cdot v)^{\ell}
((u\cdot v)\epsilon^2)^m(u^2\epsilon^2)^p(v^2\epsilon^2)^q
\,.
\end{equation}
Once we are sure that all $J_{\mu(s)}$ in \eqref{eq:def J} are traceless,
the generating function will be left unchanged by transformations  of the form $(\epsilon^{\mu})^s\to (\epsilon^{\mu})^s+(\eta^{\mu(2)})^{\ell}(\epsilon^2)^\ell(\epsilon^{\mu})^{s-2\ell}$.
We thus may use transformations of this type to effectively constraint the polarization vector to be null ($\epsilon^2=0$) without affecting the generating function of the currents,
hence without affecting the generating functions of the correlation functions either.
Thus, the only coefficients that we need to know explicitly are $f_{k,l,0,0,0}$.
The tracelessness condition $\partial_\epsilon^2f=0$ gives several relations between the coefficients appearing in \eqref{eq:series f}. 
Among those equations, we find that the $m$ dependance of the coefficients $f_{k,l,m,0,0}$ is given by
\begin{equation}
f_{k,\ell,m+1,0,0}=
-\frac{(k+1)(\ell+1)}{2(m+1)(k+\ell+m+1+\tfrac{d-2}{2})}
f_{k+1,\ell+1,m,0,0}
\,.
\end{equation}
Then, altogether with the conservation condition $\partial_\epsilon\partial_xf\vert_{u^2=v^2=0}=0$, it gives the $k$ dependence as
\begin{equation}
f_{k+1,\ell,m,0,0}
=
-\frac{(\ell+1)(\ell+m+\tfrac{d-2}{2})}{(k+1)
(k+m+\tfrac{d-2}{2})}
f_{k,\ell+1,m,0,0}
\,.
\end{equation}
Then, choosing $b_{\ell}:=f_{0,\ell,0,0,0}$, one can solve 
those two recursions and get the following expression for the 
on-shell part of the current:
\begin{equation}
f_{k,\ell,m,0,0}
=
\frac{(-1)^k}{2^m}
\frac{(k+\ell+2m)!}{k!\,\ell!\,m!}
\frac{\Gamma(k+\ell+m+\tfrac{d-2}{2})
\Gamma(\tfrac{d-2}{2})}
{\Gamma(k+m+\tfrac{d-2}{2})\Gamma(\ell+m+\tfrac{d-2}{2})}
b_{k+\ell+2m}
\,.
\end{equation}
After effectively removing $\epsilon^2$ from the generating function $J(x,\epsilon)$,
this amounts to rewrite \eqref{eq:series f} as
\begin{equation}
\label{eq:series f epsnull}
f(u,v,\epsilon)
=
\sum_{s,k}
b_s\binom{s}{k}
\frac{\Gamma(s+\tfrac{d-2}{2})\Gamma(\tfrac{d-2}{2})}
{\Gamma(k+\tfrac{d-2}{2})\Gamma(s-k+\tfrac{d-2}{2})}
(-\epsilon \cdot u)^k(\epsilon \cdot v)^{s-k}
\;.
\end{equation}
Now we choose $b_s$ to be expressed in term of a constant 
$\gamma$ (to be specified later) as
\begin{equation}
\label{eq:b_s} 
b_s=
\frac{\gamma^s}{s!\,\Gamma(s+\tfrac{d-2}{2})}
\,.
\end{equation}

We are interested in the connected correlation functions 
$\langle J_1 \cdot J_{n_0} \rangle_{\rm conn.}$, 
which descent from the contribution of several Feynman diagrams.
Since $\phi^i$ can only be contracted with $\cc{\phi}_i$, 
those contributions only differ by permutations of the currents.
As we will discuss later, this is to contrast with the real field theory, where each current has two possible contraction with the next one.
We are interested in the cyclic building block for the correlation function,
that is to say the first Wick contraction, 
that we define with the following normalisation:
\begin{align}
\CFTpa
:&=
\frac{1}{N}
\left.
\prod_{i=1}^{n_0}
f\left(\partial_{x'_i},\partial_{x_i},\epsilon_i\right)
\prod_{j=1}^{n_0}
\CFTwc{\phi^{i_j}(x_j)}{\cc{\phi}_{i_{j+1}}(x'_{j+1})}
\right\vert_{x'_k=x_k \forall k}
\\&=
\left.
\prod_{i=1}^{n_0}
f\left(\partial_{x'_i},\partial_{x_i},\epsilon_i\right)
\prod_{j=1}^{n_0}
\left\vert x_{j}-x'_{j+1}\right\vert^{2-d}
\right\vert_{x'_k=x_k \forall k}
\\\label{eq:CFTpa1}&=
\prod_{i=1}^{n_0}
f\left(-\partial_{x_{i-1,i}},\partial_{x_{ii+1}},\epsilon_i\right)
\prod_{j=1}^{n_0}
\left\vert x_{j,j+1}\right\vert^{2-d}
\;.
\end{align}
The $\frac{1}{N}$ factor has disappeared in the second line because of the internal traces.
We define $x_{ij}$ as in the previous section (now it is manifestly a 3-vector).
The rest of this section involves multiple sums that generally will be written as follows:
\begin{equation}
\sum_s\prod_{i=1}^{n}...=\sum_{s_1}...\sum_{s_n}\prod_{i=1}^{n}...
\end{equation}
This being said, we inject \eqref{eq:series f epsnull} and 
\eqref{eq:b_s} into \eqref{eq:CFTpa1} and get
\begin{align}
\nonumber&\CFTpa
\\&=
\sum_{s,k}
\prod_{i=1}^{n_0}
\frac{\gamma^{s_i}\Gamma(\tfrac{d-2}{2})}
{k_i!\,\Gamma(k_i+\tfrac{d-2}{2})(s_i-k_i)!\,
\Gamma(s_i-k_i+\tfrac{d-2}{2})}
\left(\epsilon_i\cdot\partial_{i-1,i}\right)^{k_i}
\left(\epsilon_i\cdot\partial_{i,i+1}\right)^{s_i-k_i}
\prod_{j=1}^{n_0}
\left\vert x_{j,j+1}\right\vert^{2-d}
\nonumber\\&=
\sum_{s,k}
\prod_{i=1}^{n_0}
\frac{\gamma^{s_i-k_i+k_{i+1}}\Gamma(\tfrac{d-2}{2})}
{k_{i+1}!\,\Gamma(k_{i+1}+\tfrac{d-2}{2})(s_i-k_i)!\,
\Gamma(s_i-k_i+\tfrac{d-2}{2})}
\left(\epsilon_{i+1}\cdot\partial_{i,i+1}\right)^{k_{i+1}}
\left(\epsilon_i\cdot\partial_{i,i+1}\right)^{s_i-k_i}
\left\vert x_{i,i+1}\right\vert^{2-d}
\nonumber \\ &=
\sum_{n,p}
\prod_{i=1}^{n_0}
\frac{\gamma^{n_i+p_{i}}\Gamma(\tfrac{d-2}{2})}
{p_i!\,\Gamma(p_i+\tfrac{d-2}{2})n_i!\,
\Gamma(n_i+\tfrac{d-2}{2})}
\left(\epsilon_{i+1}\cdot\partial_{i,i+1}\right)^{p_i}
\left(\epsilon_i\cdot\partial_{i,i+1}\right)^{n_i}
\left\vert x_{i,i+1}\right\vert^{2-d}
\end{align}
\begin{align}
\Leftrightarrow \; \CFTpa &=
\sum_{n,p,q}
\prod_{i=1}^{n_0}
\frac{\left(-2\right)^{n_i+q_i}\gamma^{n_i+p_{i}}}
{(n_i-p_i+q_i)!(p_i-q_i)!q_i!}
\frac{\Gamma(q_i+n_i+\tfrac{d-2}{2})}
{\Gamma(p_i+\tfrac{d-2}{2})\Gamma(n_i+\tfrac{d-2}{2})}
\nonumber\\&\qquad\times
((\epsilon_i\cdot\epsilon_{i+1})\GYisv{i}{i+1}^2)^{p_i-q_i}
(\epsilon_i\cdot\GYisv{i}{i+1})^{n_i-p_i+q_i}
(\epsilon_{i+1}\cdot\GYisv{i}{i+1})^{q_i}
\left\vert x_{i,i+1}\right\vert^{2-d}
\nonumber\\
\label{eq:CFTpa2}&=
\sum_{t,q,m}
\prod_{i=1}^{n_0}
\frac{\left(-2\right)^{t_i+q_i+m_i}\gamma^{t_i+q_i+2m_i}}
{t_i!\,q_i!\,m_i!}
\frac{\Gamma(t_i+q_i+m_i+\tfrac{d-2}{2})}
{\Gamma(t_i+m_i+\tfrac{d-2}{2})\Gamma(q_i+m_i+\tfrac{d-2}{2})}
\nonumber\\&\qquad\times
((\epsilon_i\cdot\epsilon_{i+1})\GYisv{i}{i+1}^2)^{m_i}
(\epsilon_i\cdot\GYisv{i}{i+1})^{t_i}
(\epsilon_{i+1}\cdot\GYisv{i}{i+1})^{q_i}
\left\vert x_{i,i+1}\right\vert^{2-d}\;.
\end{align}
Additionally to some index redefinition and reorganization of the 
product,
we made use of the following lemma for 2 null vectors $\epsilon_i$ and 
$\epsilon_{i+1}$ :
\begin{align}
(\epsilon_i\cdot\partial_x)^n(\epsilon_{i+1}\cdot \partial_x)^p
(x^2)^{-\tfrac{d-2}{2}}
=
&\sum_{q}
\left(-2\right)^{n+q}
\frac{n!p!}{(n-p+q)!(p-q)!q!}
\frac{\Gamma(q+n+\tfrac{d-2}{2})}{\Gamma(\tfrac{d-2}{2})}
\nonumber\\&\times
((\epsilon_i\cdot\epsilon_{i+1})\GYisv{i}{i+1}^2)^{p-q}
(\epsilon_i\cdot\GYisv{i}{i+1})^{n-p+q}
(\epsilon_{i+1}\cdot\GYisv{i}{i+1})^{q}
(x^2)^{-\tfrac{d-2}{2}}
\;.
\end{align}
The $p=0$ version can be shown recursively, then the full one comes 
from a direct application of Leibniz rule. 
We will then need
\begin{align}
k_d(t,q,m)
:&=
\frac{\Gamma(t+q+m+\tfrac{d-2}{2})}
{t!\,q!\,\Gamma(t+m+\tfrac{d-2}{2})\Gamma(q+m+\tfrac{d-2}{2})}
\nonumber\\&=
\sum_r
\frac{1}{(t-r)!\,(q-r)!\,r!\,\Gamma(r+m+\tfrac{d-2}{2})}
\;.
\end{align}
The last equality is straightforward to show when $t=0$. 
In the other cases, it is proven via the recursion :
\begin{equation}
k_d(t+1,q,m)
=
\frac{1}{t+1}\big(
k_d(t,q,m)+k_d(t,q-1,m+1)
\big)
\;.
\end{equation}
This allows to rewrite \eqref{eq:CFTpa2} as

\begin{align}
\CFTpa
&=
\sum_{t,q,m,r}
\prod_{i=1}^{n_0}
\frac{\left(-2\gamma\right)^{t_i+q_i+2m_i}}
{(t_i-r_i)!\,(q_i-r_i)!\,m_i!\,r_i!
\Gamma(r_i+m_i+\tfrac{d-2}{2})}
\nonumber\\&\qquad\times
\left(-\tfrac12(\epsilon_i\cdot\epsilon_{i+1})\GYisv{i}{i+1}^2\right)^{m_i}
(\epsilon_i\cdot\GYisv{i}{i+1})^{t_i}
(\epsilon_{i+1}\cdot\GYisv{i}{i+1})^{q_i}
\left\vert x_{i,i+1}\right\vert^{2-d}
\\&=
\sum_{a,b,c,r}
\prod_{i=1}^{n_0}
\frac{\left(-2\gamma\right)^{a_i+b_i+2c_i}}
{a_i!\,b_i!\,(c_i-r_i)!\,r_i!
\Gamma(c_i+\tfrac{d-2}{2})}
\nonumber\\&\qquad\times
\left(-\tfrac12(\epsilon_i\cdot\epsilon_{i+1})\GYisv{i}{i+1}^2\right)^{c_i-r_i}
(\epsilon_i\cdot\GYisv{i}{i+1})^{a_i+r_i}
(\epsilon_{i+1}\cdot\GYisv{i}{i+1})^{b_i+r_i}
\left\vert x_{i,i+1}\right\vert^{2-d}
\\&=
\sum_{c}
\prod_{i=1}^{n_0}
\frac{1}{c_i!\,\Gamma(c_i+\tfrac{d-2}{2})}
\exp\left(
-2\gamma(\epsilon_i+\epsilon_{i+1})\cdot\GYisv{i}{i+1}
\right)
\nonumber\\&\qquad\times
\left(-2\gamma\sqrt{
(\epsilon_i\cdot\GYisv{i}{i+1})(\epsilon_{i+1}\cdot\GYisv{i}{i+1})
-\tfrac{1}{2}((\epsilon_i\cdot\epsilon_{i+1}))\GYisv{i}{i+1}^2
}
\right)^{2c_i}
\left\vert x_{i,i+1}\right\vert^{2-d}
\,.
\end{align}
Then, as usual one defines the conformal structures as
\begin{equation}
\label{eq:def PQ CFT}
\GYciQ{i}{i-1}{i+1}=2\epsilon_i\cdot(\GYisv{i-1}{i}+\GYisv{i}{i+1})
\;,\quad
\GYciP{i}{i+1}^2=4\left(
(\epsilon_i\cdot\GYisv{i}{i+1})(\epsilon_{i+1}\cdot\GYisv{i}{i+1})
-\tfrac{1}{2}((\epsilon_i\cdot\epsilon_{i+1}))\GYisv{i}{i+1}^2
\right)
\;.
\end{equation}
In this language, we can write the final expression for the 
$n$-point conserved-current correlation functions of the 
$d$-dimensional $U(N)$ free vector model as
\begin{equation}
\label{eq:CFTpa3}
\CFTpa
=
\prod_{i=1}^{n_0}
\exp\left(-\gamma Q_i\right)
\sum_{c_i}
\frac{1}{c_i!\,\Gamma(c_i+\tfrac{d-2}{2})}
\left(\gamma P_{i,i+1}\right)^{2c_i}
\left\vert x_{i,i+1}\right\vert^{2-d}
\;.
\end{equation}

Before specifying the dimension 
in order to compare with the result in the $4$-dimensional bulk, 
let us show that this is consistent with the result conjectured in \cite{Sleight:2016dba}.
We start by rewriting \eqref{eq:CFTpa3} as a series expansion:
\begin{align}
&\CFTpa
=
\sum_{c,d}
\prod_{i=1}^{n_0}
\frac{1}{d_i!\,c_i!\,\Gamma(c_i+\tfrac{d-2}{2})}
\left(-\gamma Q_i\right)^{d_i}
\left(\gamma P_{i,i+1}\right)^{2c_i}
\left\vert x_{i,i+1}\right\vert^{2-d}
\\&=
\sum_{c,s}
\prod_{i=1}^{n_0}
\frac{1}{(s_i-c_i-c_{i-1})!\,c_i!\,\Gamma(c_i+\tfrac{d-2}{2})}
\left(-\gamma Q_i\right)^{s_i-c_i-c_{i-1}}
\left(\gamma P_{i,i+1}\right)^{2c_i}
\left\vert x_{i,i+1}\right\vert^{2-d}
\\&=
\sum_{c,s}
\prod_{i=1}^{n_0}
\frac{(-\gamma)^{s_i}}{s_i!\,c_i!\,\Gamma(c_i+\tfrac{d-2}{2})}
\partial_{Q_i}^{c_i+c_{i-1}}
\left(Q_i\right)^{s_i-c_i-c_{i-1}}
\left(P_{i,i+1}\right)^{2c_i}
\left\vert x_{i,i+1}\right\vert^{2-d}
\\&=
\sum_{s}
\prod_{i=1}^{n_0}
\frac{(-\gamma)^{s_i}}{s_i!}
\sum_{c_i}
\frac{(-)^{c_i}}{2^{2c_i}\,c_i!\,\Gamma(c_i+\tfrac{d-2}{2})}
\left(-4P_{i,i+1}^2\partial_{Q_i}\partial_{Q_{i+1}}\right)^{c_i}
\left(Q_i\right)^{s_i}
\left\vert x_{i,i+1}\right\vert^{2-d}
\\&=
\sum_{s}
\prod_{i=1}^{n_0}
\frac{(-\gamma)^{s_i}2^{\tfrac{d-2}{2}-1}}{s_i!}
\left.
(q_i)^{\tfrac12-\tfrac{d-2}{4}}
{J}_{\tfrac{d-2}{2}-1}(\sqrt{q_i})
\right\vert_{q_i=-4P_{i,i+1}^2\partial_{Q_i}\partial_{Q_{i+1}}}
\left(Q_i\right)^{s_i}
\left\vert x_{i,i+1}\right\vert^{2-d}
\;,
\end{align}
where ${J}_\alpha(x)$ is the Bessel function of first kind. 
It is now clear that the Bose symmetrisation of this result is the same as in \cite{Sleight:2016dba} up to a function of $s_i$ appearing in front of the current $J_{s_i}$.
Let us stress that this information is encoded is the weight $a_s$ appearing in \eqref{eq:def J} rather than in the normalisation $N_s$ of the current $J_s$.
Hence the freedom to fix it is not spoiled by the previous fixation \eqref{eq:b_s}  of $b_s\propto a_s\,N_s$.

Now let us go back to our three-dimensional holographic purpose. In this setup, one can use the following consequence of the duplication formula for Gamma functions:
\begin{equation}
\Gamma(x+\tfrac12)=\frac{\sqrt{\pi}(2x)!}{2^{2x}x!}
\;,
\end{equation}
and rewrite the result \eqref{eq:CFTpa3} as:
\begin{equation}
\label{eq:CFTpa4}
\CFTpa
=
\prod_{i=1}^{n_0}
\frac{1}{\sqrt{\pi}}
\exp\left(-\gamma Q_i\right)
\cos\left(2i\gamma P_{i,i+1}\right)
\left\vert x_{i,i+1}\right\vert^{2-d}
\;.
\end{equation}
We already see that if we choose $\gamma=\tfrac{i}{4}$, we recover the formula \eqref{eq:result PAb} for the type A-model,
up to global normalisation of the $n$-point function,
provided one can link the two definitions of the conformal structures $Q_i$ and $P_{i,j}\,$.
This is done by defining the polarization vector $\epsilon_i$ in terms of the polarization spinor $\GYpsu{i}$ of the previous section as follows:
\begin{equation}
\label{eq:eps(khi)}
\left(\GYpsu{i}\right)_{\alpha}\left(\GYpsb{i}\right)_{\alphadot}
=\epsilon_i^\mu\left(\sigma_{\mu}\right)_{\alpha\alphadot}
\;.
\end{equation}
It is formally a null 4-vector but 
from \eqref{eq:def K,khi,nu} 
we see that its $r$-component vanishes identically, 
making it a null vector tangent to the boundary,
as expected.

The above translation to the language of polarisation spinors 
shows that the result is the same as the one given in \cite{Gelfond:2013xt}.
We expect this to be the case in any dimension where the spinorial
language exists (e.g. $d=4$), though the verification requires a 
dictionary between the generalised space-time of \cite{Gelfond:2013xt} 
and the standard space-time where the CFT lives.

Now we want to study the O(N)-vector model. The field is now a 
real scalar, with a propagator given by
\begin{equation}
\CFTwc{\phi^{i}(x)}{\phi^{j}(y)} 
=c_2\,
\frac{\delta^{ij}}{\left\vert x-y\right\vert^{d-2}}\,
\;.
\end{equation}
Since in this case there are twice more contractions to 
consider, 
$f(u,v,\epsilon)$ will be effectively projected on its part 
that is invariant under the exchange of $u$ and $v\,$.
From \eqref{eq:series f epsnull} it is clear that this is 
equivalent to select even powers of $\epsilon\,$. Thus the 
relevant projection is
\begin{equation}
f^{O(N)}(u,v,\epsilon)
=
\tfrac{1}{2}\left(
f(u,v,\epsilon)+f(u,v,-\epsilon)
\right)
\;.
\end{equation}
From \eqref{eq:eps(khi)}, this is exactly equivalent to the 
minimal bosonic projection \eqref{eq:MB prop} of the 
propagator.

\newpage 

\section{Conclusions}
\label{sec:Conclusions}

In this paper we have examined a set of 
higher spin invariants in Vasiliev's theory, 
to the leading order in the perturbative expansion 
for curvatures, generated by 
boundary-to-bulk propagators
in the polarization spinor basis.
These invariants are given by straight,  
twisted open Wilson lines in the 
noncommutative twistor $Z$ space, with 
the most general adjoint insertions at one end.
As in noncommutative Yang--Mills theory, 
one can argue that these quantities
form a complete set of observables in 
higher spin gravity on a topologically
trivial $X$ space.
In the leading order, the observables are 
given by free (boson and fermion) CFT boundary 
correlation functions in the polarization spinor
basis, given by bounded (trigonometric) functions
times inverse powers of relative positions, 
integrable in three dimensions.

It would be interesting to extend our analysis to 
Vasiliev's D-dimensional Type A model 
\cite{Vasiliev:2003ev}, for which one can
define analogs of all the objects that we
have studied in the four-dimensional case,
and ask whether the decorated twisted open 
Wilson lines, now in the $2(D+1)$-dimensional 
phase space of the underlying conformal particle, 
reproduces the $n$-point correlation functions 
in $d=D-1$ dimensions given in \eqref{eq:CFTpa3}. 

It would also be interesting to examine 
the twisted open Wilson lines at subleading orders 
in the curvature expansion of the zero-form charges.
As the twisted, noncommutative open Wilson lines 
are invariant under the full 
higher spin gauge transformations that 
contain subleading corrections as well, 
one may speculate that these
structures contain information about 
deviations of the conformal field theory 
dual beyond the free point.
Another related issue concerns the fact 
that the zero-form charges can be evaluated 
on exact solutions to Vasiliev’s theory, 
which one may expect correspond
to states of the conformal field theory 
generated by deformations using \emph{finite sources}.
Indeed, as the exact solutions are given 
using polarization spinor bases, 
it would be interesting
to trace more carefully the observed finiteness of 
the zero-form charges \cite{Iazeolla:2011cb,Iazeolla:2012nf} 
to the mildly divergent 
nature of the correlation functions in the 
polarization spinor spaces. 

One might go further and speculate that the free energy
functional itself of Vasiliev's theory, \emph{i.e.} its
on-shell action, contains similar sub-leading corrections
related to non-linear sources coupling to 
higher-trace operators, which would thus show up as 
contact terms in the holographic correlation functions.
Such a refinement would imply that the holographically 
dual theory is actually a non-trivial three-dimensional 
field theory already in the case of the Type A Vasiliev model.
We hope to return to these issues in the near future. 

%
%
%

\section*{Acknowledgements}

We thank Carlo Iazeolla and Zhenya Skvortsov for their comments
on a preliminary version of the paper.
N.B. --- who is F.R.S.-FNRS Research Associate (Belgium) ---  
wants to thank Nicol\`o Colombo for collaboration 
at an early stage of this project as well as Fabien Buisseret for 
discussions on the Wilsonian approach to Yang-Mills theory, 
including its noncommutative extension.
The work of R.B. was supported by a PDR ``Gravity and 
extensions'' from the F.R.S.-FNRS (Belgium).
The work of P.S. is supported by Fondecyt Regular grant N$^{\rm o}$ 
1140296, Conicyt grant DPI 20140115 and UNAB internal grant DI-1382-16/R.
N.B. thanks the Galileo Galilei Institute for Theoretical 
Physics (GGI) for its hospitality and INFN for partial support 
during the completion of this work, within the program ``New 
Developments in AdS3/CFT2 Holography''.

\appendix

\section{Twisted, straight noncommutative Wilson lines}
\label{sec:thmWL}

The purpose of this appendix is to prove the central relation between observables
in Vasiliev's theory and Wilson lines 
in noncommutative Yang--Mills theory:
\begin{equation}
\label{eq:e(iMS)=W*e(iMZ)}
\starexp{i\Mu\widehat{S}} = \WLczyx{L_{2\Mu}}{Z}{Y}{x}\star
\starexp{i\Mu Z}
\;.
\end{equation}
In this section we use the convention that repeated (uncontracted) indices are totally symmetrized, 
e.g. $(Z^{\underalpha})^{\star 2}$ denotes
$\tfrac{1}{2}\left(Z^{\underalpha_1}\star Z^{\underalpha_2}
+Z^{\underalpha_2}\star Z^{\underalpha_1}\right)$.

Addressing the left hand side of 
\eqref{eq:e(iMS)=W*e(iMZ)} requires 
the use the following property of the $\star$-product:
\begin{equation}
\label{eq:[Z,f]}
\starcomm{Z_{\underalpha}}{\fhat}=
-2i\partial^Z_{\underalpha}\fhat
\;.
\end{equation}
Its successive application can be shown recursively to yield:
\begin{equation}
\label{eq:Z^m*A}
\starpuis{Z_{\underalpha}}m \star \widehat{A}_{\underalpha} = \sum_{j=0}^m \binom{m}{j} (-2i\partial^Z_{\underalpha})^j\widehat{A}_{\underalpha} \star \starpuis{Z_{\underalpha}}{m-j}
\;.
\end{equation}
Then we define the symbols $C(m,n,k_1,...,k_n)$ entering the expansion of a monomial 
in~$\widehat{S}_{\underalpha}$~as 
\begin{align}
\label{eq:doV^m}
&\starpuis{Z_{\underalpha}-2i\hat{A}_{\underalpha}}{m}
\\ &=
\sum_{n=0}^m \sum_{k_1 = 0}^{m-n} ... \sum_{k_n = 0}^{m-n-\lilsum_{i=0}^{n-1}k_i} 
(-2i)^{n+\lilsum k} C(m,n,k_1,...,k_n) \left(
\starprod_{i=1}^{n}
(\partial^Z_{\underalpha})^{k_i} \widehat{A}_{\underalpha} \right) 
\star \starpuis{Z_{\underalpha}}{\left(m-n-\lilsum k\right)}
\nonumber \;,
\end{align}
where $\sum k$ denotes $\sum_{i=1}^n k_i$ and we recall that, 
for any set of $n$ functions $\left\{\fhat_1,...,\fhat_{n}\right\}$,
the symbol $\starprod_{i=1}^{n}\fhat_i$ is defined as $\fhat_1\star...\star\fhat_{n}$ in that order.
The consistency of this expansion with \eqref{eq:Z^m*A} requires the boundary conditions: 
\begin{equation}
\label{eq:bc C}
C(m,0)=C(m,m,0,...,0)=1
\;,
\end{equation}
as well as the recurrence relations:
\begin{align}
\label{eq:rec C}
&C(m+1,n,k_1,...,k_n) 
\nonumber\\
&= C(m,n,k_1,...,k_n) + \binom{m+1-n-\sum_{i=1}^{n-1}k_i}{k_n}C(m,n-1,k_1,...,k_{n-1})
\;.
\end{align}
This can be seen by plugging \eqref{eq:Z^m*A}
into $\widehat{S}_{\underalpha}^{\star(m+1)}
=\widehat{S}_{\underalpha}^{\star m}\star\widehat{S}_{\underalpha}$.
Then defining $q:=m-n-\sum k$ allows one to reformulate the left hand side of \eqref{eq:e(iMS)=W*e(iMZ)} as:
\begin{align}
\starexp{i\Mu\widehat{S}}
&=
\sum_{n=0}^{\infty} \sum_{k_1 = 0}^{\infty} ... \sum_{k_n = 0}^{\infty}
\sum_{q=0}^{\infty} 
(2\Mu^{\underalpha})^{n+\lilsum k}(i\Mu^{\underalpha})^q
\nonumber\\[2mm]&\quad\times
\frac{C(q+n+\sum k,n,k_1,...,k_n)}{\left(q+n+\sum k\right)!}
\starprod_{i=1}^{n} \left((\partial^Z_{\underalpha})^{k_i} \widehat{A}_{\underalpha} \right) 
\star \starpuis{Z_{\underalpha}}{q}
\;.
\end{align}

On the other hand, to address the r.h.s. of \eqref{eq:e(iMS)=W*e(iMZ)} 
we recall the definition of the path-ordered exponential:
\begin{align}
&
\WLczyx{L_{2M}}{Z}{Y}{x}
=
\Pexp{2\Mu^{\underalpha} \int_0^1 \diff\sigma \widehat{A}_{\underalpha}\zyx{Z+2\sigma\Mu}{Y}{x}}
\\&=
\sum_{n=0}^{\infty} 
\left(2\Mu^{\underalpha}\right)^{n}
\int_{0}^1 \diff{\sigma_n} ... \int_{0}^{\sigma_{2}}\diff{\sigma_1}
\widehat{A}_{\underalpha}\zyx{Z+2\sigma_1\Mu}{Y}{x}\star ... \star \widehat{A}_{\underalpha}\zyx{Z+2\sigma_n\Mu}{Y}{x}
\nonumber \;.
\end{align}
Then we can write the Taylor expansion of the higher-spin connection and get:
\begin{align}
&
\WLczyx{L_{2M}}{Z}{Y}{x}
\nonumber\\&=
\sum_{n=0}^{\infty} \sum_{k_1=0}^{\infty}... \sum_{k_n=0}^{\infty}
\left(2\Mu^{\underalpha}\right)^{n+\lilsum k}
\int_0^1 \diff{\sigma_n}\frac{\sigma_n^{k_n}}{k_n!} ... \int_0^{\sigma_{2}} \diff{\sigma_1}\frac{\sigma_1^{k_1}}{k_1!}
\left(\starprod_{i=1}^{n}
\left((\partial^Z_{\underalpha})^{k_i}\widehat{A}_{\underalpha}\zyx{Z}{Y}{x}\right)\right)
\nonumber\\&=
\sum_{n=0}^{\infty}\sum_{k_1=0}^{\infty}...\sum_{k_n=0}^{\infty}
(2\Mu^{\underalpha})^{n+\lilsum k}
F(1\vert\, n,k_1,...,k_n)
\left(\starprod_{i=1}^{n}
\left((\partial^Z_{\underalpha})^{k_i}\widehat{A}_{\underalpha}\zyx{Z}{Y}{x}\right)\right)
\;,
\end{align}
Where the coefficient is given by
\begin{align}
F(\sigma\vert\, n,k_1,...,k_n) :&= 
\int_0^{\sigma} \diff{\sigma_n}\frac{\sigma_n^{k_n}}{k_n!} ... \int_0^{\sigma_{2}} \diff{\sigma_1}\frac{\sigma_1^{k_1}}{k_1!}
\nonumber \\&=
\sigma^{n+\lilsum k}
\prod_{j=1}^{n}\frac{1}{k_j!(j+\sum_{i=1}^{j}k_i)}
\;.
\end{align}
The latter form can be shown recursively. 
Upon expanding:
\begin{equation}
\starexp{i\Mu Z}
=
\sum_{q=0}^{\infty}\frac{(i\Mu Z)^{q}}{q!}
\;,
\end{equation}
we see that \eqref{eq:e(iMS)=W*e(iMZ)} holds if and only if:
\begin{align}
C(q+n+\lilsum k,n,k_1,...,k_n)
&= 
\frac{(q+n+\sum k)!}{q!} F(1\vert n,k_1,...,k_n)
\nonumber \\&= 
\frac{(q+n+\sum k)!}{q!}
\prod_{j=1}^{n}\frac{1}{k_j!(j+\sum_{i=1}^{j}k_i)}
\;.
\end{align}
It is the case, since taking this as a definition for $C(m,n,k_1,...,k_n)$ is consistent with \eqref{eq:bc C} and \eqref{eq:rec C}.
Therefore, we have proven \eqref{eq:e(iMS)=W*e(iMZ)}.

\section{Spinor notation and properties of 
the $\Sigma_i$ matrices}
\label{app:CS}

The metric of $AdS_4$ is expressed in Poincar\'e coordinates, 
so that $\text{d}s^2 = \frac{1}{r^2}\eta_{\mu\nu}\dx^\mu\dx^\nu\,$.
The Minkowski metric $\eta_{\mu\nu}$ is the one which one uses 
in this context in order to raise and lower world indices, 
as well as for computing norms of Lorentz vectors. 

The Pauli matrices plus the identity $\left(\sigma^\mu\right)_{\alpha\alphadot}$ form a basis for 
Hermitian $2\times2$ matrices. As usual, their barred 
counterpart are given by:
\begin{equation}
\left({\bar\sigma}^\mu\right)^{\alphadot\alpha}
:=
\epsilon^{\alphadot\gammadot}\,\epsilon^{\alpha\gamma}\,
\left({\sigma}^\mu\right)_{\gamma\gammadot} 
\equiv \left(\sigma^\mu\right)^{\alpha\alphadot}
\;.
\end{equation}
The main property of these matrices is
\begin{align}
\label{eq:Cl u}
\left(\sigma^\mu\right)^{\alpha\gammadot}
\left({\bar\sigma}^\nu\right)_{\gammadot}^{\phantom\gammadot\beta}
+
\left(\sigma^\nu\right)^{\alpha\gammadot}
\left({\bar\sigma}^\mu\right)_{\gammadot}^{\phantom\gammadot\beta}
&=
2\eta^{\mu\nu}\epsilon^{\alpha\beta}
\;,\\\label{eq:Cl b}
\left({\bar\sigma}^\mu\right)^{\alphadot\gamma}
\left({\sigma}^\nu\right)_{\gamma}^{\phantom\gamma\betadot}
+
\left({\bar\sigma}^\nu\right)^{\alphadot\gamma}
\left({\sigma}^\mu\right)_{\gamma}^{\phantom\gamma\betadot}
&=
2\eta^{\mu\nu}\epsilon^{\alphadot\betadot}
\;.
\end{align}
It means that they generate a Clifford algebra.
Now we omit all spinorial indices, it being understood 
that we do all contractions following the NW-SE conventions.

A Lorentz vector $A_{\mu}$ is related to two Hermitian 
$2\times2$ matrices in the following way:
\begin{equation}
A:=A^{\mu}\sigma_{\mu}
\;,\quad 
\bar{A}:=A^{\mu}{\bar\sigma}_{\mu}
\;.
\end{equation}
Then for 2 vector fields with $A_\mu(x)$ and $B_{\mu}(x)$ as respective 
component, the relations \eqref{eq:Cl u} and \eqref{eq:Cl b} become 
\begin{equation}
A\bar{B}+B\bar{A}=2A^\mu B_\mu
\;,\quad
\bar{A}B+\bar{B}A=2A^\mu B_\mu
\;.
\end{equation}
In particular:
\begin{equation}
\det{A}=A\bar{A}=A^{\mu}A_{\mu}
\;.
\end{equation}
Another useful algebraic property is:
\begin{equation}
\label{eq:flatten CS}
A(a\bar{A}+b\bar{B})B
=
a\det(A)B+b\det(B)A
=
B(a\bar{A}+b\bar{B})A
\;,\quad
\forall a,b\in\mathbb{C}
\;.
\end{equation}
Those few results are used implicitely all along the paper 
and are valid for any Hermitian matrices, 
in particular for those introduced in 
(\ref{eq:def CS app},\,\ref{eq:def CSSn}).

Now we consider a bulk point with coordinate $x^\mu_0$ 
(among which the $r$ component is denoted $r_0$) 
and various boundary points having coordinates $x^\mu_i\,$, 
$i=1,\ldots, n_0\,$. 
Their $r$ component vanishes for every boundary points.
We then define the matrices
\begin{equation}
\label{eq:def CS app}
\CSu{i}:=\sigma^{r}-2r_0\GYoisv{i}
\equiv
(\eta^{\mu r}-2r_0\GYoisv{i}^{\mu})\sigma_\mu
\;, \qquad i=1,\ldots, n_0\;.
\end{equation}
They have unit determinant and enjoy the following useful property:
\begin{equation}
\CSdm{i}{j}
:=
\det\left(\CSu{i}-\CSu{j}\right)
=
\frac{4r_0^2(\GYsv{i}{j})^2}{(\GYosv{i})^2(\GYosv{j})^2}
\;.
\end{equation}
Among the applications of \eqref{eq:flatten CS}, 
the following one is often used in our computations:
\begin{equation}
\left(\CSu{i}-\CSu{j}\right)
\left(\CSb{j}-\CSb{k}\right)
\left(\CSu{k}-\CSu{i}\right)
=
-\CSdm{i}{j}\left(\CSu{k}-\CSu{i}\right)
-\CSdm{j}{k}\left(\CSu{i}-\CSu{j}\right)
\;.
\end{equation}
One matrix inversion given in the next Appendix \ref{app:gauss} 
uses some quantities that we now define. 
Let us suppose that we have $(n+1)$ boundary points corresponding 
to the Hermitian matrices $\CSu{0},\CSu{1},...,\CSu{n}\,$.
Note that, although they are all defined in terms of the bulk 
point $x_0\,$, see \eqref{eq:def CS app}, 
the matrix $\CSu{0}$ refers to 
a boundary point that we also call $x_0\,$. This abuse of notation 
should not create confusions, hopefully. 
In other words, the boundary point corresponding to $\CSu{0}$ 
has nothing to do with the reference bulk point.
Then we define:
\begin{equation}
\label{eq:def CSSn}
\CSSnu{n}
:=
-\sum_{j=0}^{n}(-1)^j\CSu{j}
\;,\quad
\CSSniu{n}{i}
:=
\sum_{j=1}^n(1-\delta_{ij})(-1)^{j+\Theta(i,\,j)}
\CSu{j}
\;.
\end{equation}
Among the properties of the matrices $\CSSnu{n}$ and $\CSSniu{n}{i}$ 
used in Appendix \ref{app:gauss}, 
there are various applications of \eqref{eq:flatten CS}
as well as the following relations:
\begin{equation}
\CSSniu{n}{i}-\CSSniu{n}{i\pm1}=\pm(-)^{i}
\left(\CSu{i}-\CSu{i\pm1}\right)
\;,\quad
\CSSnu{n}+\CSSniu{n}{1}
=
\CSu{1}-\CSu{0}
\;,\quad 
\CSSnu{n}+\CSSniu{n}{n}
=
\CSu{n}-\CSu{0}
\;, 
\end{equation}
where the first relation is understood to hold only when both 
$i$ and $i\pm1$ take value in $\{1,...,n\}$.

\section{Inverse and determinant of the matrix $R$}
\label{app:gauss}

The results \eqref{eq:Gdmad} to \eqref{eq:Gimad last} come from various block inversions,
and the procedure is not exactly the same when $e$ and $t$ take different values.

First, let us set $e=1$.
In the two relevant cases, we decompose $R$ into the following blocks :
\begin{equation}
\label{eq:BD e=1}
R = 
\begin{pmatrix}
A_{ij} & B_{i\jbar} \\ C_{\ibar j} & D_{\ibar\jbar}
\end{pmatrix}
=
\begin{pmatrix}
(1-\Bk{i}{j})(-)^{i+j+\Theta(i,\,j)}
&
\left(\Bk{i}{\jbar}-(1-t)\delta_{i,n_0}\right)\CSu{i}
\\
\left(\Bk{\ibar}{j}-(1-t)\delta_{j,n_0}\right)\CSb{j}
&
(1-\Bk{\ibar}{\jbar})(-)^{\Theta(\ibar,\,\jbar)}
\end{pmatrix}
\;.
\end{equation}
Let us remind that in those cases, the indices takes the following values :
\begin{equation}
i\in\{1,...,n_0\}
\;,\quad 
\ibar\in\{1,...,\nbar=n_0-(1-t)\}
\;.
\end{equation}
The keypoint is that here $\nbar$ is always even, which provides $D$ with the following inverse matrix :
\begin{equation}
\inv{D}_{ij} = (1-\Bk{i}{j})(-)^{i+j+\Theta(i,\,j)}.
\end{equation}
Hence M can be decomposed as :
\begin{equation}
\begin{pmatrix}
A&B\\C&D
\end{pmatrix}
=
\begin{pmatrix}
\idmat&B\\0&D
\end{pmatrix}
\begin{pmatrix}
\tilde{A}&0\\\inv{D}C&\idmat
\end{pmatrix}
,\qquad\text{with }
\tilde{A}:=A-B\inv{D}C
\;.
\end{equation}
This allows to write its determinant and inverse matrix as :
\begin{align}
\label{eq:block det}
\det{R} &= \det(\tilde{A})\det(D)
\;,\\
\label{eq:block inv}
\inv{R} 
&=
\begin{pmatrix}
\inv{\tilde{A}}&0\\-\inv{D}C\inv{\tilde{A}}&\idmat
\end{pmatrix}
\begin{pmatrix}
\idmat&-B\inv{D}\\0&\inv{D}
\end{pmatrix}
\nonumber\\&=
\begin{pmatrix}
\inv{\tilde{A}}
&
-\inv{\tilde{A}}B\inv{D}
\\
-\inv{D}C\inv{\tilde{A}}
&
\inv{D}+\inv{D}C\inv{\tilde{A}}B\inv{D}
\end{pmatrix}
\;.
\end{align}
Let us point out the fact that since barred and unbarred indices do not run the same range, the $\delta$ symbols are summed as follows :
\begin{equation}
\sum_{i=1}^{n_0}\Bk{i}{\jbar}f(\jbar) = (1-(1-t)\delta_{i,n_0})f(i)
\;.
\end{equation}
Knowing that fact, one can easily compute :
\begin{equation}
\tilde{A}_{ij}
=
(1-\Bk{i}{j})(-)^{i+j+\Theta(i,\,j)}\CSu{i}(\CSb{i}-\CSb{j})
=
\sum_{k=1}^{n_0}
\left(\Bk{i}{k}\CSu{i}
\right)\left(
(1-\Bk{k}{j})(-)^{k+j+\Theta(k,\,j)}(\CSb{k}-\CSb{j})
\right)
\;.
\end{equation}
The purpose of this factorization is that the first block diagonal piece
is straightforward to invert and of unit determinant.
For the second piece (wich we write $X^{(n_0)}$), we proceed to a recursive block inversion,
which means computing its inverse and determinant
by decomposing it as:
\begin{equation}
X^{(n_0)}
=
\begin{pmatrix}
D'_{ij}&C'_{i}\\B'_{j}&A'
\end{pmatrix}
=
\begin{pmatrix}
\left(X^{(n_0-1)}\right)_{ij}
&
\left(X^{(n_0)}\right)_{in_0}
\\
\left(X^{(n_0)}\right)_{n_0j}
&
\left(X^{(n_0)}\right)_{n_0n_0}
\end{pmatrix}
\;.
\end{equation}
As a recursion hypothesis, we postulate the determinant and inverse of $X^{(n_0)}$ to be : 
\begin{align}
\label{eq:det Xn}
\det(X^{(n_0)}) 
&=
2^{2(n_0-2)}\prod_{i=1}^{n_0}\CSdm{i}{i+1}
\;,\\
\label{eq:inv Xn}
\inv{\left(X^{(n_0)}\right)}_{ij} 
&= 
\sum_{\eta=\pm1}\frac{1}{2\CSdm{i}{i+\eta}}
\left(-\Bk{i}{j}\eta+\Bk{i+\eta}{j}\xi_{i,i+\eta}\right)
(\CSu{i}-\CSu{i+\eta})
\;,
\end{align}
where we recall:
\begin{align}
\xi_{i,i+\eta}
&=
-\eta + t\;\delta_{i,n_0}\,(\eta+1)+t\;\delta_{i+\eta,n_0}\,(\eta-1)
\;,\\
\CSdm{i}{j}
:&=
\det\left(\CSu{i}-\CSu{j}\right)
=
\frac{4r_0^2(\GYsv{i}{j})^2}{(\GYosv{i})^2(\GYosv{j})^2}
\;.
\end{align}
When proceeding to the recursive step, it is important to note 
that $0$ is identified with $(n_0-1)$ in the expression of $\inv{\left(X^{(n_0-1)}\right)}_{ij}$, 
implying $(1)-1=(n_0-1)$ and $(n_0-1)+1=1\,$.
Knowing that, one finds:
\begin{equation}
\tilde{A}'
=
\frac{2}{\CSdm{n_0-1}{n_0}}(\CSb{n_0-1}-\CSb{n_0})(\CSu{n_0-1}-\CSu{1})(\CSb{n_0}-\CSb{1})
\;.
\end{equation}
Then the algebra of $\CSu{i}$ matrices (see appendix \ref{app:CS}) gives its inverse and determinant as :
\begin{align}
\det\tilde{A}'
&=
\frac{4\CSdm{1}{n_0-1}\CSdm{1}{n_0}}{\CSdm{n_0-1}{n_0}}
\;,\\
\inv{\left(\tilde{A}'\right)}
&=
\frac{1}{2\CSdm{1}{n_0}\CSdm{1}{n_0-1}}
(\CSu{n_0}-\CSu{1})(\CSb{n_0-1}-\CSb{1})(\CSu{n_0-1}-\CSu{n_0})
\;.
\end{align}
Then the straightforward application of \eqref{eq:block det} and of \eqref{eq:block inv} allows first to confirm \eqref{eq:det Xn} and \eqref{eq:inv Xn},
then to find equations \eqref{eq:Gdmad} to \eqref{eq:Gimad last}.

In the cases where $t=1$, we use the following block decomposition:
\begin{equation}
\label{eq:BD t=1}
R = 
\begin{pmatrix}
D_{ij} & C_{i\jbar} \\ B_{\ibar j} & A_{\ibar\jbar}
\end{pmatrix}
=
\begin{pmatrix}
(1-\Bk{i}{j})(-)^{i+j+\Theta(i,\,j)}
&
\left(\Bk{i}{\jbar}-(1-e)(-1)^{i+n_0}\delta_{j,n_0}\right)\CSu{j}
\\
\left(\Bk{\ibar}{j}-(1-e)(-1)^{j+n_0}\delta_{i,n_0}\right)\CSb{i}
&
(1-\Bk{\ibar}{\jbar})(-1)^{\Theta(\ibar,\,\jbar)}
\end{pmatrix}
\;.
\end{equation}
Here $n=n_0-(1-e)$ is the one to be always even, and the same exact method can be applied to this cases.

In the last case, the one where $e=t=0$, both barred and unbarred indices run an odd number of values. Therefore, we take a different approach and start from the following block decomposition :
\begin{equation}
R
= 
\begin{pmatrix}
B_{ij} & A_{i\jbar} \\ D_{\ibar j} & C_{\ibar\jbar}
\end{pmatrix}
= 
\begin{pmatrix}
A_{i\jbar} & B_{ij} \\ C_{\ibar\jbar} & D_{\ibar j}
\end{pmatrix}
\begin{pmatrix}
0 & \idmat \\ \idmat & 0
\end{pmatrix}
\;.
\end{equation}
We decide in the rest of this appendix to write $\CSu{n_0}$ as $\CSu{0}$, in order to do recursions on $n:=n_0-1$ without having to change notations between two recursive steps.
As explained in Appendix \ref{app:CS}, the boundary point corresponding to $\CSu{0}$ 
has nothing to do with the reference bulk point.
The matrix to invert in this case is :
\begin{equation}
N
:=
\begin{pmatrix}
A_{i\jbar} & B_{ij} \\ C_{\ibar\jbar} & D_{\ibar j}
\end{pmatrix}
=
\begin{pmatrix}
\Bk{i}{\jbar}\CSu{i}+(-1)^{i}\CSu{0}
&
(1-\Bk{i}{j})(-1)^{i+j+\Theta(i,\,j)}
\\
(1-\Bk{\ibar}{\jbar})(-1)^{\Theta(\ibar,\,\jbar)}
&
\Bk{\ibar}{j}\CSb{i}+(-1)^{j}\CSb{0}
\end{pmatrix}
\;.
\end{equation}
The first step is to recursively block invert $D$ and get :
\begin{align}
\inv{D}_{i\jbar} 
&=
\Bk{i}{j}\CSu{i}+\frac{(-1)^{j}}{\CSSnd{n}}\CSu{i}\CSSnb{n}\CSu{j}
\;,\\
\det{D}&=\CSSnd{n}
\;,
\end{align}
where $\CSSnu{n}$ is defined by \eqref{eq:def CSSn}.
After some algebra, the next step gives :
\begin{equation}
\tilde{A}_{i\jbar}
=
-\frac{(-1)^i}{\CSSnd{n}}
\left(\CSSnu{n}+(-1)^{\Theta(i,\,j)}\CSSniu{n}{i}\right)
\CSSnb{n}
\left(\CSSnu{n}-(-1)^{\Theta(i,\,j)}\CSSniu{n}{j}\right)
\;,
\end{equation}
where $\CSSniu{n}{i}$ is defined by \eqref{eq:def CSSn}.
It is good to note that the diagonal piece does not depend on the convention we choose for $\Theta(i,\,i)$.
Then the recursive block inversion of $X^{(q)}$ (the upper left $q\times q$ block of $\tilde{A}$ gives:
\begin{align}
\inv{\left(X^{(q)}\right)}_{\ibar j}
&=
\sum_{\substack{\eta=\pm1\\0\neq i+\eta\neq q+1}}
\frac{\eta(-1)^i}{2\CSdm{i}{i+\eta}}
\nonumber \\
&\left(
\frac{\delta_{ij}}{\det{\left(\CSSnu{n}-\CSSniu{n}{i}\right)}}
\left(\CSSnb{n}-\CSSnib{n}{i}\right)
\left(\CSSnu{n}-\CSSniu{n}{i+\eta}\right)
+\delta_{i+\eta,j}\right)
\left(\CSSnib{n}{i}-\CSSnib{n}{i+\eta}\right)
\nonumber\\&\quad
+\frac{\delta_{ij1}}{\det{\left(\CSSnu{n}-\CSSniu{n}{1}\right)}\det{\left(\CSSnu{n}+\CSSniu{n}{1}\right)}}
\left(\CSSnb{n}+\CSSnib{n}{1}\right)
\CSSnu{n}
\left(\CSSnb{n}-\CSSnib{n}{1}\right)
\;,\\
\det X^{(q)}
&=
\frac{2^{2(q-1)}}{\CSSnd{n}\det{\left(\CSSnu{n}+\CSSniu{n}{1}\right)}\det{\left(\CSSnu{n}-\CSSniu{n}{q}\right)}}
\;.
\end{align}
Then the straightforward application of \eqref{eq:block det},\eqref{eq:block inv} and of the properties of $\CSu{}$
allow once again to recover the equations \eqref{eq:Gdmad} to \eqref{eq:Gimad last}.

\newpage
\providecommand{\href}[2]{#2}\begingroup\raggedright\endgroup

\end{document}